\documentclass[12pt]{article}
\pdfoutput=1

\usepackage{graphicx}
\usepackage{psfrag}
\usepackage{enumerate}
\usepackage{soul}
\usepackage{subfig}
\usepackage{jcappub}
\usepackage{latexsym}
\usepackage{natbib}
\usepackage{dsfont}

\newcommand{\be}{\begin{equation}}
\newcommand{\ee}{\end{equation}}

\def\[{\left [}
\def\]{\right ]}
\def\({\left (}
\def\){\right )}
\def\p{\partial}

\def\r2{\sqrt{2}}

\def\O{{\mathcal O}}

 \def\simleq{\; \raise0.3ex\hbox{$<$\kern-0.75em
      \raise-1.1ex\hbox{$\sim$}}\; }
   \def\simgeq{\; \raise0.3ex\hbox{$>$\kern-0.75em
      \raise-1.1ex\hbox{$\sim$}}\; }

\newcommand{\rmd}{\mathrm{d}}
\newcommand{\de}{\partial}

\newcommand{\gam}{\gamma}
\newcommand{\Gam}{\Gamma}
\newcommand{\del}{\delta}

\newcommand{\lam}{\lambda}

\newcommand{\sig}{\sigma}

\newcommand{\nab}{\nabla}

\renewcommand{\k}{\vec{k}}
\newcommand{\avg}[1]{\langle #1 \rangle}

\newcommand{\bbibitem}[1]{\bibitem{#1}\marginpar{#1}}

\newcommand{\figref}[1]{Fig. \ref{#1}}
\newcommand{\secref}[1]{Sec. \ref{#1}}
\newcommand{\tableref}[1]{Table \ref{#1}}
\newcommand{\appref}[1]{Appendix \ref{#1}}

\def\Label#1{\label{#1}%
  \smash{\hbox to0pt{\raise1ex\hbox{\tiny[#1]}\hss}}}
\def\noLabels{\let\Label=\label}
\def\nobbibitem{\let\bbibitem=\bibitem}

\begin{document}

\noLabels
\nobbibitem

\DeclareGraphicsExtensions{.pdf,.png,.gif,.jpg,.eps}

\title{Unwinding Inflation}
\author[\bigstar]{Guido D'Amico,}
\author[\bigstar]{Roberto Gobbetti,}
\author[\bigstar,\heartsuit]{Matthew Kleban,}
\author[\bigstar]{and Marjorie Schillo}

\emailAdd{gda2@nyu.edu, rg1509@nyu.edu, mk161@nyu.edu, mls604@nyu.edu}

\affiliation[\bigstar]{\it Center for Cosmology and Particle Physics, 
New York University, New York, NY  }
\affiliation[\heartsuit]{\it Institute for Advanced Study, 
Princeton, NJ  }

\abstract{Higher-form flux that extends in all 3+1 dimensions of spacetime is a source of positive vacuum energy that can drive meta-stable eternal inflation.   If the flux also threads compact extra dimensions, the spontaneous nucleation of a  bubble of  brane charged under the flux can trigger a classical cascade that steadily unwinds many units of flux, gradually decreasing the vacuum energy while inflating the bubble, until the cascade ends in the self-annihilation of the brane into radiation.
With an initial number of flux quanta $Q_{0} \simgeq N$, this can result in  $N$ efolds of inflationary expansion while producing a scale-invariant spectrum of adiabatic density perturbations with amplitude and tilt consistent with observation. The power spectrum has an oscillatory component  that does not decay away during inflation, relatively large tensor power, and interesting non-Gaussianities.
Unwinding inflation fits naturally into the string landscape, and our preliminary conclusion is that consistency with observation 
can be attained without fine-tuning the string parameters.  The initial conditions necessary for the unwinding phase are produced automatically by bubble formation, so long as the critical radius of the bubble is smaller than at least one of the compact dimensions threaded by flux.

}

\maketitle

\begin{center}
\begin{table}[t]
\begin{tabular}{| l | l |}
	\hline
	\multicolumn{2}{|c|}{\bf  \it Dramatis Person\ae} \\
	\hline 
	$D$ & Total dimension of space-time \\
	$F,p$ & $F$ is a $p+2$-form field strength under which a $p$-brane is charged \\
	$Q (Q_0)$ & Number of $F$ flux quanta (prior to bubble formation) \\
	$\cal M$ & A compact ($D-4$)-manifold \\
	$H, N$ & Hubble constant of the 4D space-time, number of efolds of inflationary expansion  \\
	$z_b, z$ & Radius of the brane in the compact dimensions,  effective 4D inflaton field\\
	$v , \gamma, \chi, \sigma$ & Velocity $v=\dot z$, Lorentz factor, rapidity, and tension of the brane \\
	$c_s$ & Speed of sound for perturbations $\delta z$, \emph{cf.}~\secref{subspert} \\
	$l$ & Circumference of compact directions threaded by $F$  in which the brane expands \\
	$d$ & Circumference of compact dimensions transverse to $F$ or wrapped by the brane \\
	$\zeta$  &  FRW curvature perturbation, \emph{cf.}~\secref{protosec} \\
	$\mathcal{P}_\zeta(k)$ & Dimensionless power spectrum of $\zeta$ \eqref{PSdef} \\
				$n (\rho_s)$ & Number (energy) density of strings produced per brane/anti-brane collision \\
		$\lambda$ & Time averaged friction due to string production, \eqref{lambda}. \\
	$\bar f$ & Time averaged pressure due to string production, see~\eqref{bkg1} \\
	\hline
 \end{tabular}
\end{table}
 \end{center}

\section{Introduction}

Many theories, including string theory~\cite{Susskind:2003kw}, predict the existence of  metastable phases with positive vacuum energy.   Such phases drive eternal inflation and are expected to exponentially dominate the global volume of the universe.  If so, our Hubble volume  must be contained inside a pocket or bubble embedded in the parent false vacuum.  To be consistent with observational constraints on spatial curvature, the radius of  our bubble must be at least 10 times larger than the present-day Hubble length, meaning it must have undergone $N \simgeq 60$ efolds of inflationary expansion \emph{after} it formed.

In this paper we present the details of ``unwinding inflation'' \cite{D'Amico:2012sz}, a novel mechanism by which an eternally inflating metastable false vacuum can transition via charged brane bubble formation to flux discharge cascade \cite{Kleban:2011cs} that mimics slow-roll inflation.  The vacuum energy is initially only slightly reduced by the formation of the bubble, but then steadily ``unwinds'' over time. Inflation ends and reheating occurs with the self-annihilation of the brane into radiation once most or all vacuum energy is discharged.  Without fine-tuning of either the parameters or the initial conditions, this phase can drive 60 or more efolds of  expansion, solving the curvature problem and generating a scale-invariant spectrum of perturbations.  

Unwinding inflation occurs when the vacuum energy of the parent phase is at least partially supplied by electric-type flux that extends in the 3+1 dimensions of spacetime, and in addition wraps at least one compact extra dimension (hence, a five or higher form flux).  Under these circumstances, a brane bubble that reduces the flux by one unit of the brane's charge can nucleate~\cite{Brown:1988kg}.  If the bubble is localized on the compact space, it will expand, rapidly reaching relativistic velocities, wrap around the compact dimension, collide with itself and initiate a flux cascade that repeatedly discharges the flux one unit at a time~\cite{Kleban:2011cs}.  The effective four-dimensional inflaton scalar  $z$---the radius of the bubble in the compact dimensions---increases steadily with time, as in conventional slow-roll models.
 If the size of the extra dimension is $l$ and the Hubble constant is $H$, roughly   $(H l)^{-1}$ units of flux will be discharged per Hubble time.  We believe compactifications with $H l \simgeq 1$ are difficult to achieve, and therefore to attain $N$ efolds of inflation probably requires $Q_{0} \simgeq N$, where $Q_{0}$ is the number of initial flux units.  

A heuristic analogy is that the flux is a rubber sheet that wraps multiple times around some  compact cycles, as well as extending in the 3+1 large dimensions of spacetime.  The nucleation of the brane bubble is the spontaneous appearance of a spherical hole (bounded by brane) in one layer of the sheet.  The  total energy of the brane when it appears is equal to the total energy cut out of the sheet.  Once it appears, the tension of the sheet pulls on the hole and causes it to expand in all the directions  the sheet extends in.  As the hole expands around the compact cycle and overlaps itself, it unwinds more and more layers of the sheet.

This mechanism uses ingredients (branes and flux) found in all string compactifications, and as such fits naturally into the string theory landscape~\cite{Susskind:2003kw}. The cosmological constant problem---the requirement that inflation ends at or close to zero vacuum energy---is solved in theories with sufficiently large numbers of eternally inflating phases \cite{Weinberg:1987dv}.  String or M-theory compactified on a manifold with length scale $\O(10)$ times larger than the fundamental length allows for a more than sufficient number \cite{Bousso:2000xa}, and also allows $Q_{0} \sim 100$ as required for unwinding inflation.  The two hierarchically different scales involved in bubble formation followed by slow roll that appear unnatural from the low-energy point of view  \cite{Linde:1998iw} (the thinness of the bubble wall versus the flatness of the inflationary plateau) arise naturally: the bubble wall is a $D$-brane and therefore very thin, while the flatness of the inflationary plateau follows if $Q_0 \simgeq N$.

Unwinding inflation has the virtue that it sets up its own initial conditions---eternally inflating false vacuum states are exceptionally powerful attractors~\cite{Brown:2011ry}, and unwinding inflation initiates spontaneously from them.  At the same time, this model reheats homogeneously and isotropically and explains how the bubble can become 14 Gyrs in size, thereby solving the problems that derailed Alan Guth's original model of ``old'' inflation \cite{Guth:1980zm}.  Unlike in old inflation, the rate of bubble nucleation can be arbitrarily small---a bubble will nucleate eventually, and only one is needed.

\subsection{Observational consequences}
 Several observational tests of the eternally inflating multiverse have been proposed, among them searching for the effects of cosmic bubble collisions (see \cite{Kleban:2011pg} for a recent review) and measuring non-zero spatial curvature \cite{Kleban:2012ph, Guth:2012ww}.  The weak point in these tests is the sensitivity to the duration of slow-roll inflation after the bubble forms.  Long inflation in the bubble erases all signatures of such ``initial state of the bubble" effects with exponential efficiency.  While there are reasons to believe inflation was short \cite{Freivogel:2005vv}, that conclusion is weakened by the lack of understanding of the origin of the slow-roll phase.

 By contrast, unwinding inflation predicts a set of characteristic features that do not inflate away.  Fundamentally, this is because the compact dimensions remain stable and small, but nevertheless play a continuous and key role during inflation.  Brane collisions occur periodically as the bubble expands around the compact dimensions and intersects itself, producing open and possibly closed strings.  This string production contributes an oscillatory component to the power spectrum of perturbations.  The inflaton potential arises primarily from the background flux, but it too has an oscillatory component due to brane-anti-brane interactions in the compact directions.  
 
The presence of Lorentz invariant (Dirac-Born-Infeld) kinetic terms for the brane  and the possible presence of additional light scalars (describing the transverse position of the brane in the compact dimensions) lead to non-Gaussianity.  Unlike most previous attempts to realize inflation in string theory, the scale of inflation is high, and unwinding inflation naturally predicts tensor modes with an amplitude that can be observed in the near future.  

The exact details of our observational predictions are necessarily tentative, because we do not yet have a realization in a fully stabilized string compactification, and by the same token do not know the geometry of the compactified manifold.  We will attempt to elucidate to what extent our predictions are generic and independent of the details of the compactification, and where they may break down.  At least in principle, confirmation of the predictions of this model could provide evidence for the eternally inflating multiverse and the presence of cosmic bubble collisions (albeit in a compact dimension), probe the geometry of the extra dimensions, and provide observational support for string theory.
 
\subsection{Relation to previous work} 
A short, self-contained description of this model can be found in \cite{D'Amico:2012sz}. Discharge of higher-form flux by branes was first considered in  \cite{Brown:1988kg}, and flux discharge cascades in \cite{Kleban:2011cs}.   The idea of using a bubble collision in a compact dimension to reheat homogeneously and isotropically was first proposed in~\cite{Brown:2008ea}, and the possibility of scalar cascades was mentioned in \cite{Easther:2009ft,Giblin:2010bd}. Fluctuations of bubble walls  were considered in \cite{Garriga:2001qn, Garriga:1991tb,Adams:1989su}.  Another model of inflation that uses compact extra dimensions to extend the field range is \cite{Silverstein:2008sg}, related 4D effective field theories were studied in \cite{Kaloper:2008fb, Kaloper:2011jz, Dubovsky:2011tu} and the idea of using a relativistic brane for inflation with a Dirac-Born-Infeld (DBI) action was proposed in \cite{Silverstein:2003hf, Alishahiha:2004eh}.  The effects of particle production on the spectrum of perturbations was considered in \cite{Berera:1995ie, Green:2009ds, LopezNacir:2011kk}.  Various models have been considered in the past that utilize the attractive potential between a brane and anti-brane for  inflation (\emph{e.g.}~\cite{Dvali:1998pa}).  Such models typically need warping to make the potential flat enough (\emph{e.g.}~\cite{Baumann:2007ah}).  Unwinding inflation does not, because the inflaton potential arises instead from background flux (brane/anti-brane interactions are a small, periodic perturbation in the potential).  It also evades the exponential rarity of post-tunneling slow-roll  described in \cite{Yang:2012jf}, because  the tunneling creates the conditions necessary for slow roll.  A model using multiple quantum tunnelings in place of slow roll (in contrast to unwinding inflation, which is classical after an initial bubble nucleation) is \cite{Freese:2004vs}.  After this paper was completed and \cite{D'Amico:2012sz} appeared, \cite{Shlaer:2012by} was posted, which overlaps with \cite{Kleban:2011cs, D'Amico:2012sz}, and this work.

\section{Background and basic mechanism}

Higher-form fluxes that extend in the 3+1 large dimensions of spacetime and in any number of compact extra dimensions contribute to the 4D components of the stress tensor like vacuum energy.  
In string theory (or any other theory with several extra dimensions) the different ways of threading flux through compact cycles gives rise to a large landscape of metastable vacua.  With the mild conditions that the compact dimensions be  larger than the string length by an $\O(1)$ factor and that there are $\O(100)$ distinct compact cycles, the number of possible vacua can greatly exceed $10^{120}$, thus providing the necessary conditions for an anthropic solution to the cosmological constant problem~\cite{Weinberg:1987dv, Bousso:2000xa}.

Typical vacua of this type have a large vacuum energy and inflate very rapidly, and are therefore expected to dominate the overall volume of the universe.  Bubbles of other phases form inside these rapidly inflating regions via a variety of phase transitions such as the tunneling of metastabilized scalar moduli from a minimum and the discharge of a unit of flux by the nucleation of a spherical bubble of brane~\cite{Brown:1988kg}.  Both transitions produce bubbles containing FRW cosmologies, and additional inflation is necessary to solve the curvature problem.  

Our mechanism occurs when a bubble of brane (that initially discharges one brane charge unit of flux) appears and is smaller than any of the compact directions it extends in (namely, the directions the flux threads).  Under these conditions there is a  ``flux discharge cascade" \cite{Kleban:2011cs}, where a $(p+2)$-form electric flux threading  at least one compact dimension can ``unwind'', repeatedly discharging in a cascade triggered by the quantum nucleation of a bubble of charged brane, and hence steadily decreasing the effective four dimensional vacuum energy.

\begin{figure}
\begin{center}
	{\includegraphics[width=.6\textwidth]{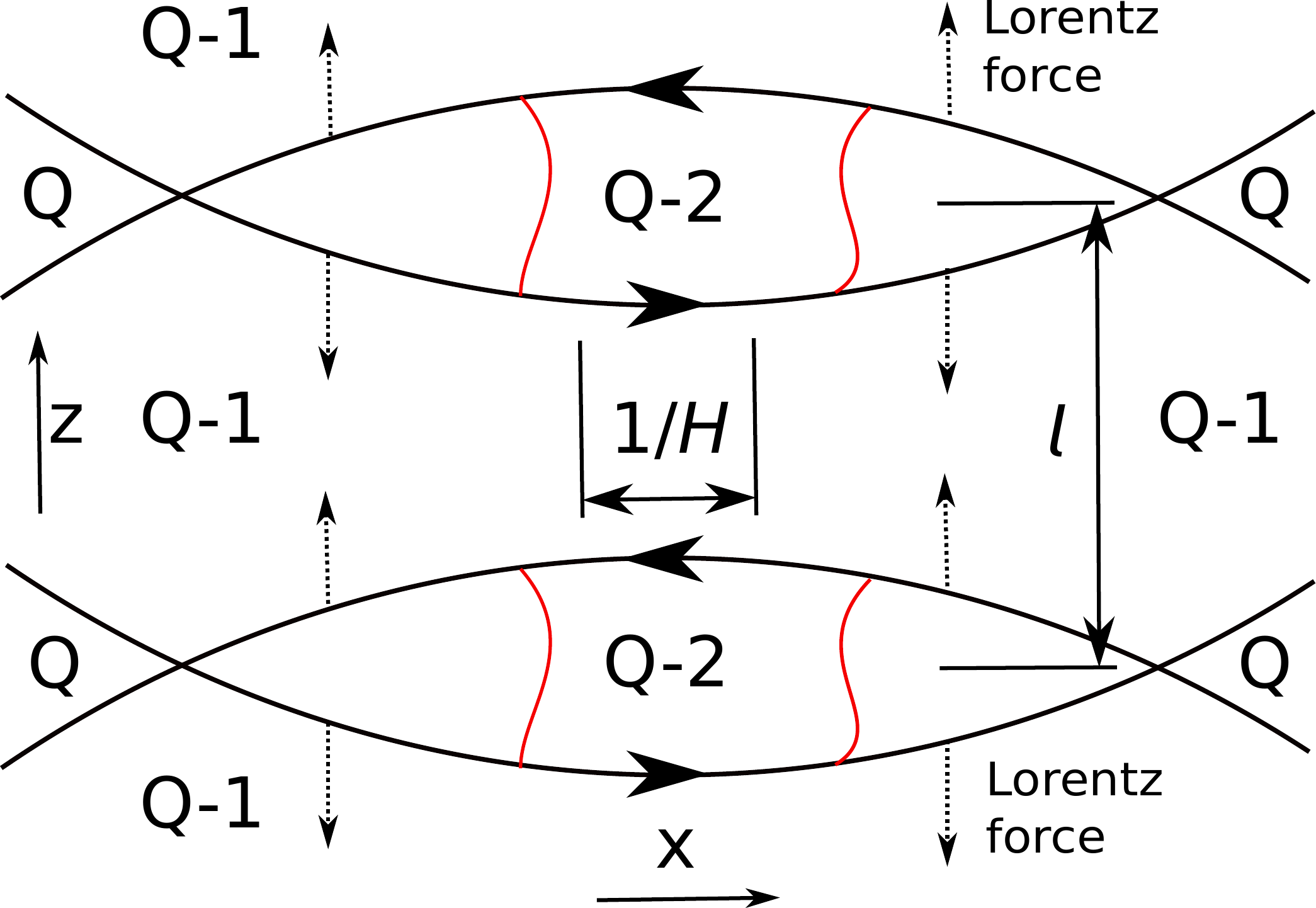}} \\
\end{center}
\caption{\label{prl} The mechanism of flux discharge cascade on $dS_{4}\times S_{1}$, where $z \simeq z+l$ is the coordinate on the $S_{1}$. The amount of flux is indicated by $Q, Q-1, \ldots$, while the dashed arrows represent the direction of the electric force and the velocity of the branes.  The squiggly lines indicate strings stretched between sections of brane; their mass depends on the separation and changes with time.  The figure is not to scale; usually $l <1/H, R$ where $R$ is the radius of curvature of the branes. }
\end{figure}

In the case of a top ($D$) form flux,  the brane is co-dimension 1 (a domain wall) in the space.  In that case the flux everywhere inside and outside the brane bubble is constant, but the interior flux is reduced relative to the exterior by one unit of the brane's charge.  The non-zero  flux at the bubble wall exerts a force on it, causing it to expand in all directions after it appears.  The bubble expands freely in the 3+1 dimensions of spacetime, but in the compact dimension(s) it wraps around and collides with itself.  Ignoring self-interactions for a moment, it passes through itself, and in the overlap region discharges the flux by two units.  This region expands and, after a second wrap, forms a region with three units discharged, etc. (\figref{prl}).

\begin{figure}
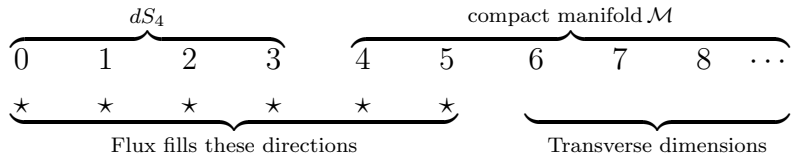

$$
\rlap{$\underbrace{\overbrace{{0\atop \star} \qquad  {1\atop \star}\qquad  {2\atop \star}\qquad  {3\atop \star}}^{dS_4}\qquad {4\atop \star}\qquad{5\atop \star}}_{\textrm{Flux fills these directions}}$} \phantom{{0\atop \star} \qquad  {1\atop \star}\qquad  {2\atop \star}\qquad  {3\atop \star}} \qquad \overbrace{\phantom{{4\atop \star}\qquad{5\atop \star}}\qquad\underbrace{{6 \atop } \qquad {7 \atop } \qquad {8 \atop } \quad {\cdots \atop }}_{\textrm{Transverse dimensions}}}^{\textrm{compact manifold} \, \mathcal{M}}
$$
\caption{The setup for unwinding inflation in the case $p=4$.}\label{schem}
\end{figure}

The necessary ingredients for unwinding inflation are (\figref{schem}):
\begin{itemize}
\item A $(p+2)$-form field strength $F$ with $p \geq 3$, and $p$-branes  that are electrically charged under $F$.
\item A $D=4+q$ dimensional spacetime  $dS_{4} \times \mathcal{M}_q$,\footnote{One could generalize this and consider a warped product, but we will not do so here.}
 where $dS_4$ is 4D de Sitter spacetime and $\mathcal{M}_{q}$ is a stabilized compact $q$-manifold with  $q \geq p-2 \geq 1$.
\item $Q_0 \gg 1$ units of $F$ flux threading $dS_{4}$ and a $p-2$ cycle in $\mathcal{M}_q$, supplying the $dS_4$ vacuum energy.
\end{itemize}

An observer located at a point in the spacetime would encounter a series of brane walls that sweep across her location at regular intervals, and a flux that is constant except when a wall crosses her position, after which it decreases by one unit.  An observer unable to resolve distance or time scales of order the size of $\mathcal{M}$ would simply observe a steadily decreasing flux.  Because the flux contributes positively to the effective vacuum energy of the 4D spacetime, during the cascade there is a gradual decrease in the Hubble constant of the de Sitter---just as in ordinary slow-roll inflation.  After part or all of the flux is discharged, the remaining effective 4D vacuum energy can be positive, negative, or zero, depending  on the stabilization mechanism and any additional fluxes or vacuum energy.  For our purposes, we will assume the  4D vacuum energy at the end of the cascade is close to zero, or at the value that will result in nearly zero vacuum energy after GUT or standard model phase transitions that occur later in the evolution of the universe.

The reduction in flux during the cascade can also lead to a change in the geometry of $\mathcal{M}$, typically reducing its overall volume. However if the flux is not the primary element that stabilizes $\mathcal{M}$ this change is small, and we will neglect it (we will comment more on this in \secref{casimir}).

\subsection{Prototype model} \label{protosec}

The simplest version of unwinding inflation has $D=5$ and $p=3$.  In this version the initial spacetime is $dS_4 \times S_1$, where the vacuum energy of the $dS_4$ is supplied in part by $Q_0$ units of initial $F_5$ flux.  Stabilizing the $S_1$ requires an additional ingredient, for example the Casimir energy of several bosonic and fermionic fields \cite{ArkaniHamed:2007gg}.  In \secref{casimir} we will describe stabilization in more detail, but for now will simply assume the $S_1$ is stable.  The objects charged under $F_5$ are $p=3$-branes, with charge $e$ and tension $\sigma$.  With $Q$ units of flux, one has $F_5 = Q e$.  In some cases, higher dimensional versions of unwinding inflation can be dimensionally reduced to this 5D version.

\paragraph{Instanton:}

The nucleation of brane bubbles that discharge the flux is governed by a solution to the Euclidean signature equations of motion~\cite{Brown:1988kg}. The Euclidean signature version of the spacetime has metric 

\be \label{eq:EucMet}
	ds_E^2=H^{-2}\left(d\xi^2 +\sin^2{\xi}d\Omega_{3}^{2}\right) + dz^2 ,
\ee
where $z\simeq z+l$ is the coordinate on  $S_1$ and $d\Omega_{3}^{2}$ is a 3-sphere.  

Typically, the dominant instanton for decay of a false vacuum state has the maximal symmetry possible.  We will assume the brane is thin, and that the instanton depends only on $\xi$ and $z$ in accord with the symmetry of the initial state (so that the bubble is spherical in the $dS_4$ directions).  We are interested in the case where the maximum size $\Delta z$ of the instanton in the $z$ direction satisfies $\Delta z < l$, so  the periodic boundary conditions do not affect the solution (at least in the thin-wall limit).  Finally, when the initial number of flux units $Q_0 \gg 1$  the gravitational backreaction of a single bubble is small and can be ignored.

With these assumptions the instanton solution is fully characterized by the location of the wall $z=\pm z_b(\xi)$, where $\pm$ refers to two symmetric halves  and we have chosen $z=\xi=0$ as the center of the bubble.  To find $z_b(\xi)$, one should minimize the action 
$S_E = - \kappa\int_{V}\sqrt{g_E} + \sigma\int_{\partial V}\sqrt{g_{E(induced)}},$
where $ \kappa$ is the difference in energy density on the two sides of the wall, $\sigma$ is the tension of the wall, and $V$ is the volume enclosed by the bubble.  

Parametrizing the position of the wall by $z=\pm z_b(\xi)$, the Euclidean action is
\be
\label{eq:bkgaction}
\begin{split}
	S_E =& H^{-4} \int dz \int d\Omega_3 d\xi \sin^3{\xi}
		\Bigg\{ -\kappa \Theta(z+z_b) \Theta(-z+z_b) \\
	&\qquad + 2\sigma  \delta(z-z_b) \[1+H^{2}\(\frac{dz}{d\xi}\)^2\]^{1/2} \Bigg\}  \\
	=& 4 \pi^2 H^{-4} \int d\xi \sin^3{\xi} \left\{ - \kappa z_b(\xi)
	+ \sigma \[ 1+H^{2}\(\frac{dz_b(\xi)}{d\xi}\)^2 \]^{1/2} \right\}.
\end{split}
\ee

Extremizing this action results in equations of motion that can be solved analytically for $dz_b/d \xi$, with the integration constants fixed by the requirements of finite action and smoothness:
\be
	\frac{dz_b}{d\xi} = \frac{i(8+9\cos\xi-\cos3\xi)}{H\sqrt{-(8+9\cos\xi-\cos3\xi)^2+144(\sigma H/\kappa)^2\sin^6\xi}}
\ee
This solution describes a bubble with spherical topology in five dimensions.  The shape is oblate; the $z$ direction is distinguished from the Euclidean de Sitter directions because it is flat, and the coordinate extent in $z$ is somewhat less than it would be in 5D Euclidean space.

We can determine the Lorentzian signature evolution of the bubble after it nucleates through the analytic continuation $\xi \to i H t, d \Omega_{3} \to i dH_{3}$.   In flat space, the bubble would expand and accelerate indefinitely with constant proper acceleration, reaching a gamma factor $\gamma = z/2 R$ after expanding by a distance $z$~\cite{Kleban:2011cs}.  Instead, because of Hubble friction the wall approaches an asymptotic velocity $v<1$ in the $z$-direction  \cite{Brown:2008ea}
\be \label{velocity}
\lim_{t \rightarrow \infty} \frac{dz_b}{dt} \equiv v = \frac{1}{\sqrt{1+\(3\sigma H/\kappa\)^2}} =  \frac{1}{\sqrt{1+\(3 R_{0} H/4\)^2}},
\ee
where $R_{0}\equiv 4\sigma / \kappa$ would be the radius of the bubble in five dimensional flat space ($H=0$).   Additional sources of friction (for instance from string production due to collisions) may reduce $v$ even further. 

Because their charges are opposite, it is useful to think of the half of the bubble at $z=+z_{b}$ as brane, and $z=-z_{b}$ as  anti-brane.  Unwinding inflation eventually ends when the brane annihilates with an image anti-brane.
For various reasons detailed below, the brane position $z_b$ will become inhomogeneous with perturbations $\delta z_b(\vec x, t)$.  Assuming that the unperturbed brane will annihilate with its image after $Q_{t}$ units of flux are discharged,  reheating occurs when 
$$z_b(t)+\delta z_b(\vec x, t)=v t + \delta z_b = Q_t l/2.$$
Solving for $t$ yields 
$$t=Q_t l/2 v-\delta z_{b}/v \equiv t_{0} + \delta t.$$  
The curvature of this hypersurface is $a(t_{0} + \delta t) = a(t_{0}) + \dot a(t_{0} )\delta t,$ and so the curvature perturbation is
$$\zeta = \delta a/a = H \delta t = H \delta z_b/v = H \delta z_b/\dot z_b,$$
where $H$ and $\dot z_b$ are evaluated at horizon crossing as usual.

\paragraph{Effective action:}

After formation of the bubble, the brane/flux action in Lorentzian signature is
 \be  \label{braneac}
	S = H^{-3} \int dz\int d H_{3} \, dt  \, \sinh^3(H t) \(-2 \sigma \delta(z-z_b)\sqrt{1-(\partial z_b)^2} - {F_{5}^2 \over 2 \cdot 5!} \) 
	\ee
where $dH_{3} = \sinh^{2} \rho \, d\rho \, d\Omega_{2}$ is the measure on a unit hyperboloid.  Gauss' law \cite{Brown:1988kg} requires that the flux changes across the brane by one unit of the brane charge: 
$$
\frac{F_5^2}{5!} = \mu^5 Q^2 =  {\mu^5} \left(Q_0 + \sum_{j=-\infty}^{\infty}\left[ \Theta(z-z_b+j l)-\Theta(z+z_b+j l) \right]\right)^2.
$$
Here $\mu^{5/2}$ is the charge of the brane, $Q_0$ is the number of flux units prior to the bubble nucleation, and the sum arises due to the periodicity of $z$.  From this, one can see that the jump in vacuum energy across a brane separating $Q$ from $Q-1$ units of flux is $\kappa(Q) = \mu^5(Q-1/2)$.

As usual for cosmic bubbles that nucleate due to first-order phase transitions, the bubble's walls and interior are naturally described using the negatively curved spatial slicing of \eqref{braneac}.  In these coordinates, the energy density inside the bubble is homogeneous and isotropic.  The spatial curvature of the bubble universe is  $\Omega_k = (a H)^{-2} \approx e^{-2N}$ after $N$ efolds of inflation, and today would be $\Omega_k \sim e^{-2(N-N_*)}$, where $N_* \sim 60$ and $N$ is the total number of efolds of inflation after the bubble forms.  Since observation constrains $\Omega_k \simleq .01$ \cite{Komatsu:2010fb}, unwinding inflation requires $N \simgeq N_*+ 3$.

After a few efolds of expansion, the radius of curvature of the bubble $R(t) \sim R_0 e^{H t}$ stretches to super horizon scales and $\Omega_{k}$ is exponentially small.  From then on during inflation, to a good approximation the bubble can be treated as a flat, parallel brane/anti-brane pair separated by a distance $2 z_b$ in the compact dimension, and the open slicing used in \eqref{braneac} can be replaced with the flat slicing (see \figref{prl}).  We will neglect corrections due to spatial curvature for the remainder of this paper.

\begin{figure}
\begin{center}$
\begin{array}{c c }
	\includegraphics[width=0.5\textwidth]{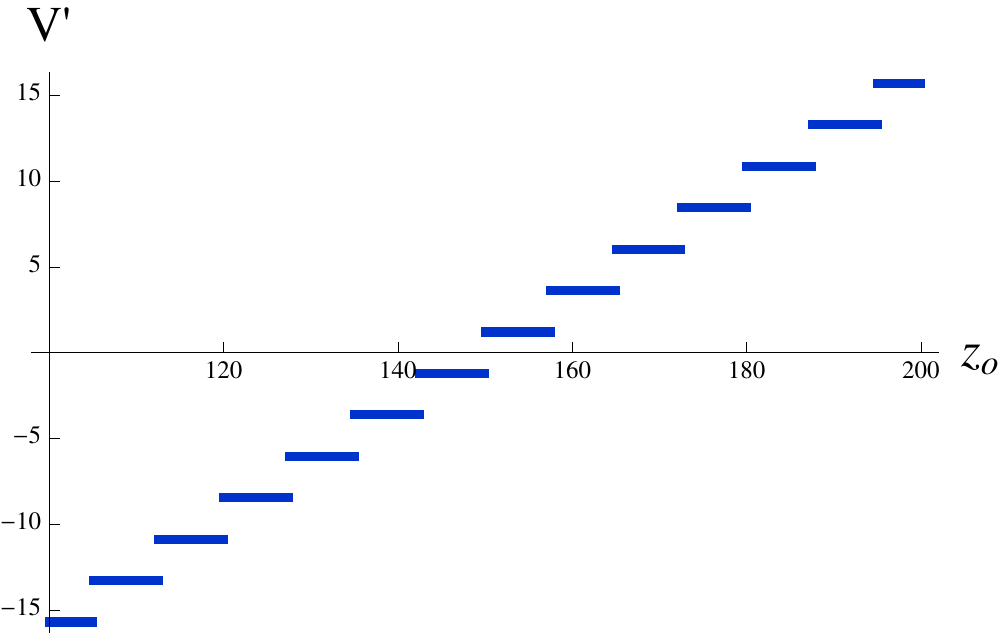} & \includegraphics[width=0.5\textwidth]{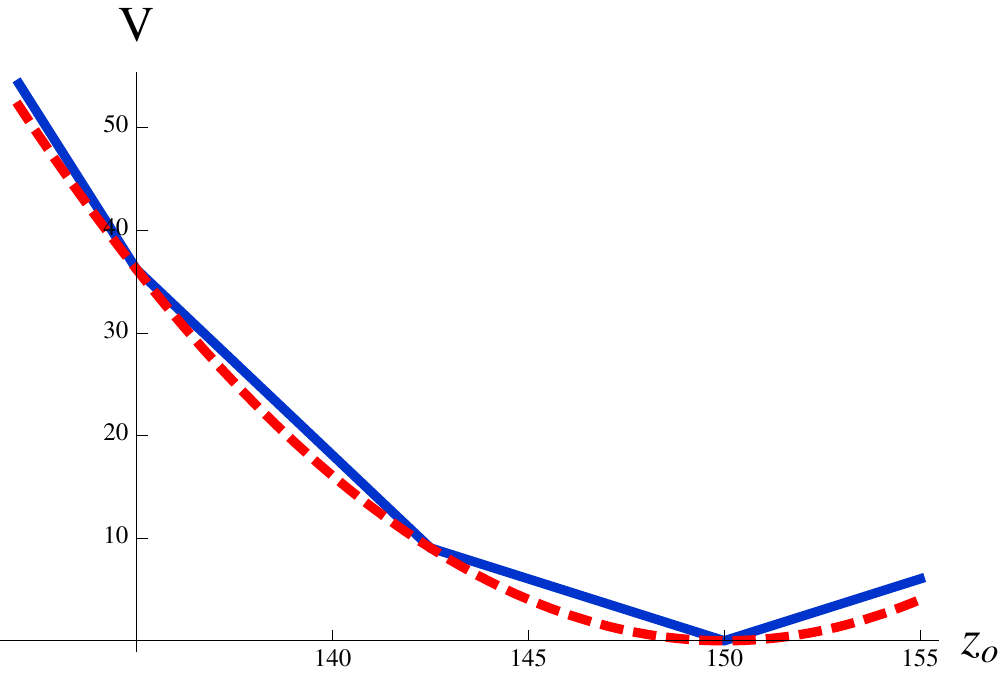} \\
\end{array}$
\end{center}
\caption{\label{Vppfig} Left panel:  The pressure $V'$ from \eqref{eq:Vpp}, constant between collisions and changing in steps at each collision.  Right panel:  The potential $V$ around the zero flux minimum; the smooth line is the quadratic approximation.  }
\end{figure}

The action \eqref{braneac} can be integrated over $z$ to obtain a 4D effective action:
\be \label{protac}
S = \int dt d^3 x e^{3 H t} \( -2 \sigma \sqrt{1-(\partial_t z_b)^2+e^{-2 H t} (\partial_{\vec x} z_b)^2}  - V(z_b) \),
\ee
where we have approximated the open slicing in \eqref{braneac} with flat slices, and $V(z_b)$ is a piecewise-linear interpolation of a quadratic that is most easily expressed in terms of its derivative (see \figref{Vppfig}): 
\be \label{eq:Vpp}
\frac{dV}{dz_b} \equiv V'(z_b)= - 2 \mu^5\left(Q_0 - \frac{1}{2} - \left[ \frac{2 z_b}{l} \right] \right),
\ee
where $\left[ ... \right]$ denotes  integer part.

This expression has a very simple physical interpretation:  $-V'(z_b)$ is the pressure on the brane, which is proportional to the (integer) number of flux units at its location.  Every time $ \left[ {2 z_b /  l} \right]$ increases by one, an additional collision has occurred and the flux (and hence the pressure $V'$) is reduced by one additional unit of the brane charge $\mu^{5}$.

\section{Four dimensional effective description}\label{4deffec}

In the previous section we derived a 4D effective action \eqref{protac} describing the simplest version of unwinding inflation.   As we will see, other compactification geometries  produce 4D actions that differ in some details, but all of them share certain features in common. In this section we will derive the cosmological predictions of unwinding inflation in terms of a few parameters of the 4D action.

The effective 4D dynamics of unwinding inflation can be described by the following ingredients.

\begin{itemize}
\item  The inflaton scalar field $z_{b}$ (corresponding to the radius of the brane bubble in the compact dimensions that are threaded by flux).\footnote{It is important to note that $z_b$ is not a fundamental scalar---it is a collective coordinate that does not exist prior to bubble nucleation.  Its potential $V(z_b)$ does not in general have false minima or a tunneling transition.}
\item $D-p-2$ additional scalars $\vec b$ (corresponding to the position of the brane in the compact directions perpendicular to the flux).  
\item Dirac-Born-Infeld  kinetic terms for all the scalars, normalized by the tension of the brane $\sigma$ and possibly functions of the fields.
\item A potential $V(z_{b})$ (arising from $Q$ units of  electric flux that force the bubble to expand, as well as from the interactions of the charge of the brane  with its images  in the compact directions).
\item A collection of extra degrees of freedom that are produced periodically in bursts when the branes collide.
\end{itemize}
The bubble nucleation sets up initial conditions in which all fields are homogeneous and isotropic on negatively curved (open universe) slices, and $V(z_{b}) \sim Q_{0}^{2}$ is relatively large
All cosmological consequences of this model can  be determined from the 4D description.  At the level of detail we will present here, {the relevant quantities} are $\sigma$,  $V$,  the average density of produced strings/particles/gravitons  $\rho_s$, and the number of  scalars along with  information regarding how they influence the time of reheating and modulate the kinetic terms.

In general, the reduction to four dimensions produces a Kaluza-Klein-type tower of modes with masses $m \sim \O(1/l)$, arising from excitations of the brane in the compact directions it extends in.  Since we consider only compactifications with $H l < 1$, these modes are heavier than the Hubble scale and do not affect the dynamics of inflation significantly. 
The 4D effective action for $z$ (for clarity from here forward we drop the subscript on $z_b$) is:
\be \label{eq:effecach}
S = -\int dt d\vec x e^{3 H t} \left\{  2 \sigma h(z) \sqrt{1-(\partial_t z)^2+e^{-2 H t} (\partial_{\vec x} z)^2}  + V(z)\right\}
\ee
(c.f.  \eqref{protac}, \eqref{s2action}).

The potential $V$ has contributions both from the unwinding flux and from other sources, such as the higher dimensional cosmological constant, other fluxes, Casimir energies, etc.  Assuming the extra dimensions remain stable throughout the cascade, these latter contributions are constant.  By contrast, the energy density due to the unwinding flux scales as $Q^2$ and therefore depends on $z$ and decreases during inflation.  As we mentioned above, the 4D vacuum energy must be tuned so as to solve the cosmological constant problem.  This means the constant term is a small contribution relative to the $\sim Q^2$ piece, and we will neglect it.  

During unwinding inflation when $Q \gg 1$,  the kinetic energy $2 \sigma h \gamma$ arising from the first term in \eqref{eq:effecach} is small compared to $V$.  This must be the case at least early on, because the brane nucleation conserves energy, and therefore the energy in the brane tension equals the energy released by discharging only one unit of flux.  Furthermore, Hubble friction limits the Lorentz factor $\gamma$, and so it is only when most or all of the flux has discharged that the kinetic energy becomes important.   Therefore, during much of the cascade the 4D Einstein equations are $H^2 \sim V$, as usual in models of slow-roll inflation.

A probe brane moving on a warped product manifold (where the radius of the $dS_4$ depends on position in the compact dimensions) gives rise to an action similar to \eqref{eq:effecach} (see \emph{e.g.} \cite{Silverstein:2003hf}), but we are not considering warping here.  Instead, in uwinding inflation the function $h(z)$ arises because $z$ is the radius of the bubble expanding in a (possibly curved) compact manifold. We give some examples of compactifications that produce non-trivial $h(z)$ in \appref{S2app} and \appref{torapp}, but for simplicity in the remainder of this section we will focus on the case where $h(z)=1$:
\be \label{effecac}
S = -\int dt d\vec x e^{3 H t} \left\{ 2 \sigma \sqrt{1-(\partial_t z)^2+e^{-2 H t} (\partial_{\vec x} z)^2}  + V(z)\right\}.
\ee

The action \eqref{effecac} is incomplete in that it only describes the radius of the bubble, and -- in the string theory context -- ignores the other open string degrees of freedom, as well as the coupling between open and closed strings (apart from the coupling to the background flux, which is taken into account  in $V$).  
The massless scalar that describes the center-of-mass position of the bubble in the compact directions does not contribute to curvature perturbations \cite{Garriga:2001qn}, because its value does not affect the time of reheating (at least at lowest order).  More importantly, strings stretched between the brane and its images have a mass that depend on $z$  (see \figref{prl}).  Because $z$ is changing in time these open  strings are produced when the brane scatters \cite{Bachas:1995kx} off its images in the compact dimensions, in a process very similar to that of \cite{Green:2009ds}.  Furthermore, closed strings may be produced by Bremsstrahlung.  Here, rather than explicitly including these modes, we will incorporate this effect by including the average amount of particle production per collision $\rho_s$ (see \cite{Green:2009ds,LopezNacir:2011kk}).  

\subsection{Background evolution} \label{bkgnd evo}
The potential energy is the energy in the flux: $V \propto F^2 \propto Q^2$, averaged over the compact directions.  The specific form  $V(z)$ depends on the compact geometry, but in all cases it decreases with increasing $z$ as $Q$ discharges, and includes an oscillating component arising from interactions of the brane with itself as it wraps the compact dimensions ({\emph c.f.} \figref{Vppfig}, \figref{fig:Vsphere}).   In several examples\footnote{When the cycle the flux threads is $S_n$ for any $n$, or when it is $S_n \times {\mathcal N}$ with the size of $\mathcal{N}$ small enough that the brane bubble wraps it rather than expanding in it.}, $V(z) \sim Q^2 \sim (Q_0 -  z/ l)^2+$ small oscillations, where $l$ is the typical lengthscale of the compact cycle the flux threads. The amplitude of the oscillations is suppressed by at least $1/Q$, since it arises from interactions of the brane with itself rather than the $Q$ units of background flux.  These examples are similar to $m^{2} \phi^{2}$ inflation, although with a DBI kinetic term and extra degrees of freedom.

\begin{figure}
\begin{center}$
\begin{array}{ccc}
	\includegraphics[angle=0, width=0.32\textwidth]{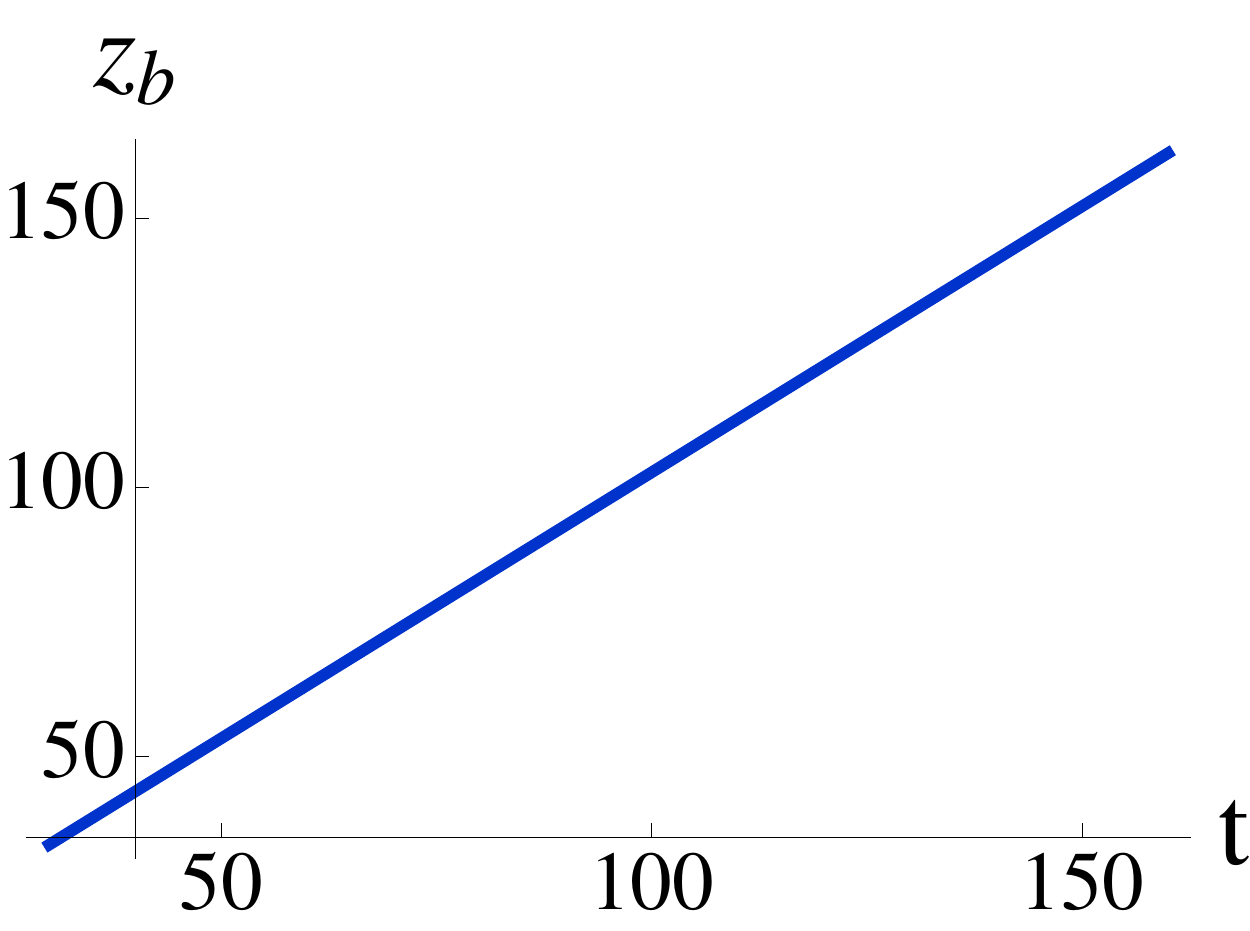} & \includegraphics[angle=0, width=0.32\textwidth]{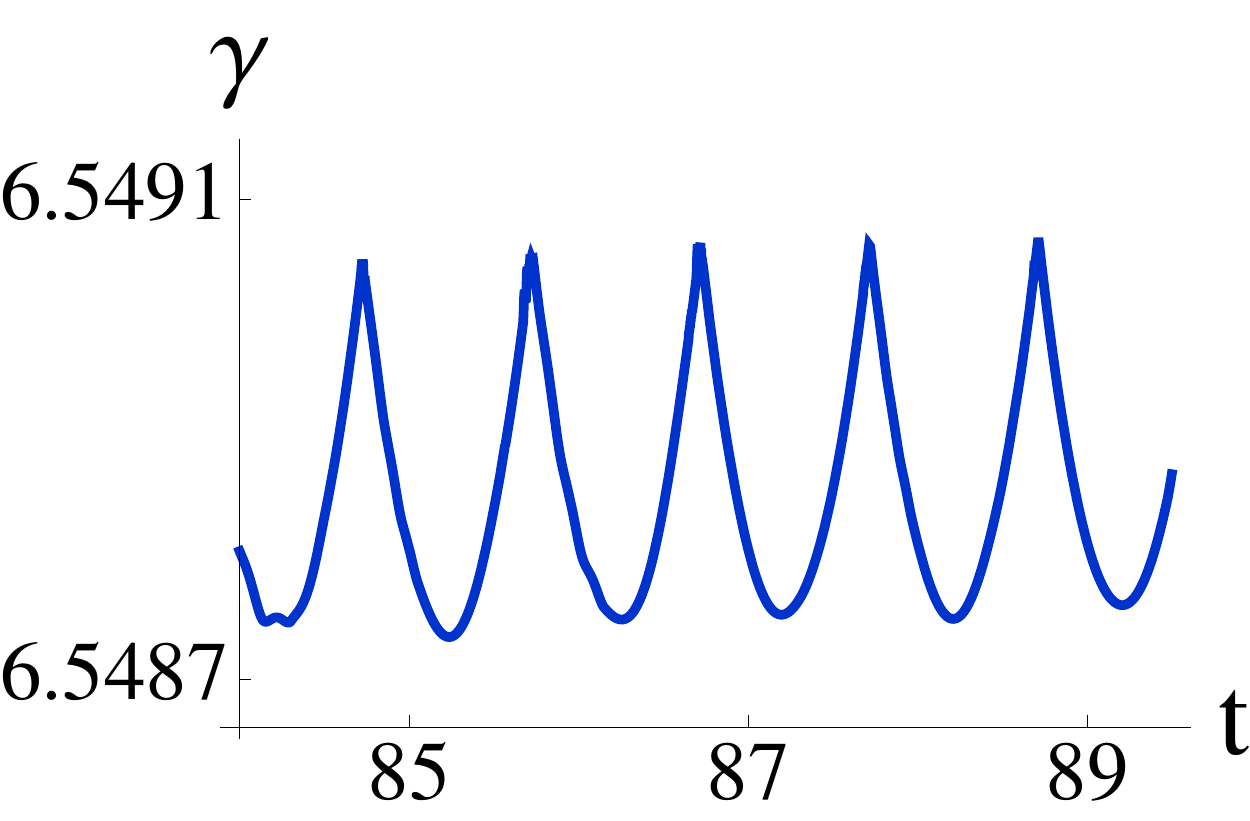} & \includegraphics[angle=0, width=0.32\textwidth]{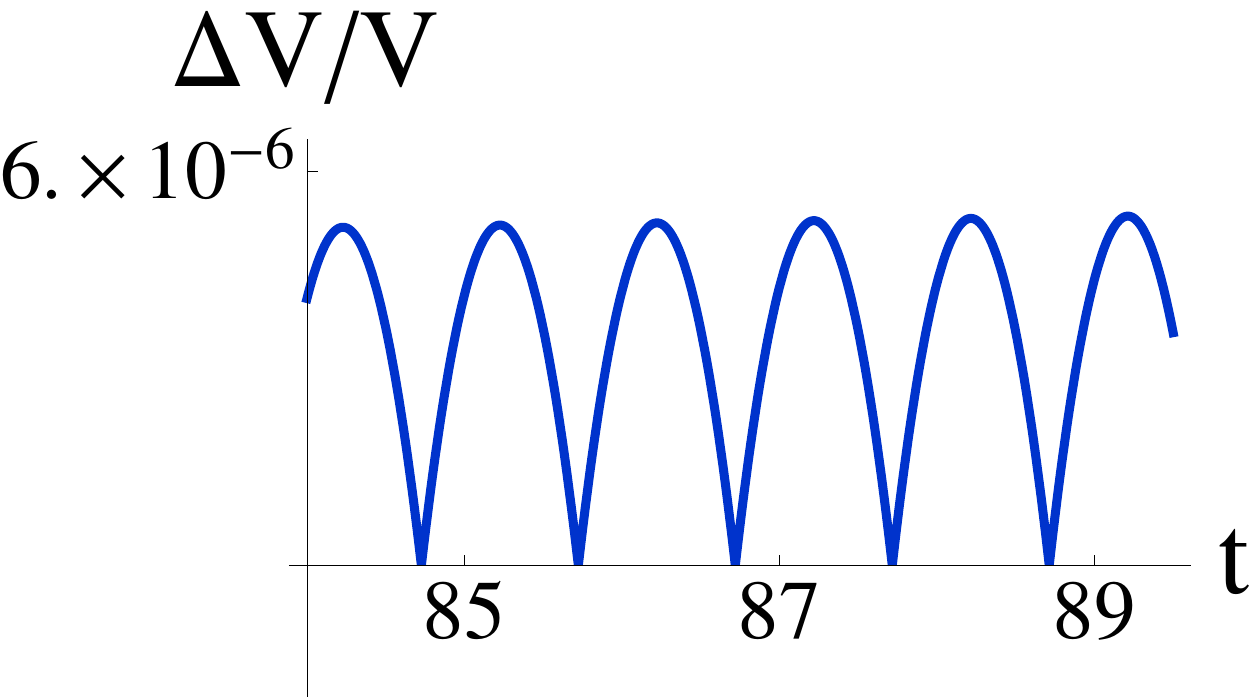} \\
\end{array}$
\end{center}
\caption{\label{fig:num} From left to right: The brane position $z$,  the Lorentz factor $\gamma$, the oscillations around the smooth approximation of the potential, zoomed in to the region corresponding to  the CMB quadrupole.
All plots use the set of parameters: $g_s=.01$, $l=19.7/m_s$, $d=2/m_s$, $Q_0=400$, for a wrapped $(p=4)$-brane expanding on an $S_1$, see  \secref{stringsec}.  As is apparent from the plots the oscillations in $\gamma$ and $V$ are very small in this case, but the power spectrum may have larger oscillations depending on the degree of string production.}  
\end{figure}

Generically the collision or scattering of the brane with its images is inelastic due to the creation of particles or strings.  This particle creation will affect the background equation of motion \eqref{bkg1} because it converts some of the kinetic energy of the brane into particles, therefore acting as a source of friction.  However, in most regimes of interest in unwinding inflation this additional friction term is subdominant to Hubble friction.  

Ignoring the effects of particle production for now, the background (homogeneous) equation of motion from \eqref{effecac} is 
\be \label{bkgeq}
	\ddot{z} +\frac{3H}{\gamma^2}\dot{z}+\frac{V'}{2 \sigma\gamma^3}=0.
\ee
 The combination of the force due to $V'$, the DBI kinetic term, and Hubble friction results in $z \sim v t$ with the slow-roll velocity $v$ nearly constant, with $1-v \ll 1$ in most cases.  The small oscillations in $V$ do not strongly affect the background evolution (see~\figref{fig:num}).

As a check on our formalism, for the ``prototype" model where $V' = 2 \kappa = 2 \mu^5(Q-1/2)$, $v=\dot z$ obtained from \eqref{bkgeq} with $\ddot z=0$ agrees exactly with the result derived from the instanton \eqref{velocity}.

\subsection{Perturbations} \label{subspert}

There are two sources of perturbations (both scalar and tensor) in this model.   First, the finite temperature of de Sitter space leads to tensor and scalar fluctuations in the standard way.  As we will see, the power spectrum of these perturbations depends only on the Hubble constant $H$ during inflation, the effective brane tension $\sigma$, and the velocity $\dot z$ (and $h(z)$ when it is not constant).  Second, the collisions of the brane bubble with itself as it wraps around the compact dimensions will produce particles and/or strings.  Determining the rate at which this happens requires a model (see \appref{annulusapp}), but the effect on the power spectrum of perturbations depends at first approximation only on the energy density produced as a function of $z$ and its time derivatives.  Particle production is a random process, and (at least assuming the theory is weakly coupled) particle/string production events should be Poisson distributed at separations larger than the Compton wavelength of the produced particles.  This assumption determines the statistics of fluctuations in the rate of particle production, and therefore the statistics of the resulting fluctuations in $z$.

The brane collisions could also give rise to gravitons, either Bremsstrahlung from scattering or via the decay of some other produced particle.  It is possible for the contribution of tensors produced this way to exceed those produced by standard de Sitter fluctuations \cite{Senatore:2011sp}, but we will defer investigating that possibility to later work.

\paragraph{De Sitter perturbations:} We begin by analyzing the model in the absence of any string or particle production.
With this simplification, the power spectrum for theories of the form \eqref{effecac} can be computed using standard methods~\cite{Garriga:1999vw}.
The late time expression for $\delta{z}$ is very simple:
\be
	\del z_k \to {H \over 2 \sqrt{\sigma k^3}} \, .
\ee
As usual, the curvature perturbation is related to perturbations in $z$ by $\zeta = H \del z/\dot{z}$.  Defining the power spectrum $\mathcal{P}_\zeta$ by
\be \label{PSdef}
	\langle \zeta(k) \zeta(k') \rangle = 2\pi^2\frac{\mathcal{P}_\zeta(k)}{k^3}\delta^3(k-k'),
\ee
we find 
\be
\label{eq:noSspectrum}
	\mathcal{P}_\zeta = \frac{H^4}{8 \pi^2 \sigma v^2} \, .
\ee
Incorporating non-trivial $h(z)$ (see \eqref{eq:effecach}) multiplies~\eqref{eq:noSspectrum} by $1/h(z)$, but this is reliable only when $h(z)$ and the speed of sound $c_s$ vary slowly.  In the case of unwinding inflation on $S_2$ for example, this adiabaticity assumption does not hold (see \appref{S2app}).
 
 Because observations determine $\mathcal{P}_\zeta\sim 2 \times 10^{-9}$, the Hubble parameter $H$ in unwinding inflation satisfies $H \sim 0.02 \, \sigma^{1/4}$.  Hence if the brane tension $\sigma$ is close to the string or Planck scale (as it is in string theory), unwinding inflation is a high-scale model of inflation.

The speed of sound---which is $c_s = 1/\gamma$ for~\eqref{eq:effecach} and \eqref{effecac}---cancels out of the power spectrum \eqref{eq:noSspectrum}, although as we will see it contributes to non-Gaussianity.  This cancellation can be understood most simply as follows.  The power spectrum can be written as $\mathcal{P}_\zeta \sim G_N^2 V^2/(c_s T)$, where $G_N$ is the 4D Newton constant, $V$ is the potential energy and $T$ is the kinetic energy.  In our case, $T \sim \gamma \sigma = \sigma/c_s$, so the speed of sound cancels.

\paragraph{Tensor power:}  The tensor power spectrum due to  de Sitter perturbations with action \eqref{effecac} is simply~\cite{Garriga:1999vw}
\be
\label{tensor}
	\mathcal{P}_{h} = \frac{16 G_{N} H^2}{\pi} \, ,
\ee
therefore the tensor-to-scalar ratio is
\be
\label{rvalue}
r=\frac{128 \pi G_{N} \sigma v^{2}}{H^{2}} \, .
\ee
As we will see in Tab.~\ref{parameters}, in models derived from string theory, $r$ is potentially observable in the near future.

\paragraph{Oscillations:}  The power spectrum \eqref{eq:noSspectrum} oscillates due to the oscillations in $V \sim H^2$.  For the ``prototype" model \eqref{protac}, the amplitude of these oscillations relative to the average is very small, $\sim Q^{-2}$ (\figref{fig:num}).  For the $S_2$ (discussed in \appref{S2app}), the oscillations in the $V$ are $\sim 1/Q$.

\paragraph{Tilt:} Because $v \sim 1$ and assuming $\sigma$ is constant,  the tilt of the scalar spectrum $n_s-1 \equiv d \ln \mathcal{P}/d \ln k$  arises from the $H^4$ term in \eqref{eq:noSspectrum}:
\be \label{tilt}
n_s-1 \approx 4 \frac{\dot H}{H} \frac{dt}{d \ln k} \approx 4 \frac{\dot H}{H^2} \, .
\ee
During unwinding inflation $H \sim \sqrt{V} \sim Q$, and $\dot{Q}$ is constant at least when $V'(z) \sim z$.  In this case, 
$$n_s -1 \approx -2/N_* \approx -0.033,
$$ 
where $N_* = \int H dt = \int dH (H/{\dot H} ) \approx -H^2/(2 \dot H) $ is the number of efolds from the time the quadrupole mode crossed the horizon during inflation to the end of inflation.  In a model with a high reheat temperature $N_* \sim 60$, and  the tilt \eqref{tilt} is   consistent with observation \cite{Komatsu:2010fb}.

 Ordinarily $V''$ contributes to the tilt because it is a mass term for the perturbation, and variations in the speed of sound contribute in the form $\dot c_s/H c_s$.  In our case these two contributions exactly cancel to first order in slow roll, as is evident from \eqref{tilt} and can be checked using eq. (35) of \cite{Garriga:1999vw}.

A potentially significant correction to \eqref{tilt} could arise from changes in $\sigma$ during inflation.  As we discuss below, if realized in string theory the brane is likely to be wrapping a compact cycle, in which case $\sigma$ depends on  the volume of that cycle.  If the volume of the compact cycle changes significantly during inflation, $\sigma$ will change as well, therefore tilting the spectrum.  However at least in toy models (see \secref{casimir}), the size of the compact cycle changes only slightly as the flux is discharged, making a small or negligible contribution to the tilt.

\subsection{Effects of particle or string production} \label{ppsec}

Each time the brane collides with itself, some fraction of its kinetic energy will be converted into particles or strings.  The particles produced by any given collision will dilute away exponentially on Hubble time scales, but because collisions happen at least once per Hubble time, the time-averaged density is not  necessarily small.  

Here, we will include the effects of   particle production without referring to the underlying model.  Our treatment is in many ways parallel to \cite{Green:2009ds, LopezNacir:2011kk}.  Specifically, the field theory model considered in \cite{Green:2009ds} is the non-relativistic limit of the sector of our model that describes the coupling of $z$ to the modes of the open strings that are massless when the brane and anti-brane coincide.  The work of \cite{ LopezNacir:2011kk} considers dissipation in more generality.  Our methods differ from these works in several ways, one being that we do not assume that dissipative effects are the dominant source of perturbations or friction in the background equation of motion.  Instead, we will derive a general result that applies to both those cases and situations where de Sitter perturbations dominate.

We will assume that the produced particles are massive and weakly interacting, so that the number density $n_c$ produced at each collision redshifts as $a^{-3}$ (this assumption can easily be relaxed), and that they are created instantaneously at times $t_{i}$ when the branes collide.  The energy density in the produced particles is
\be \label{rhos}
\rho_{s} = \sum_{i} m_c(z, z_i, \dot z_i, ...) \, n_c(\dot z_{i}, \ddot z_{i}, ...) \, e^{-3 H (t-t_{i})} \,  \Theta(t-t_{i}),
\ee
where $m_c$ is the mass of the string or particle produced by the collision,  and $z_{i}, \dot z_{i}$, etc. are the position, velocity, etc. of the brane at the collision times $t_i$.  In string theory as well as in the field theory considered in \cite{Green:2009ds}, the  mass of the produced particles/strings grows linearly with $z-z_i$, and we will assume $m_c \propto z-z_i$, although this can be generalized without difficulty.\footnote{The reader might be concerned that this assumption implies that the particles are massless when created, but that the modes of stretched open strings do not necessarily have this property, depending on their oscillator mode.  However, in string theory the string production occurs in a time $\Delta t \sim 1/m_s \ll l, 1/H$, and the typical mass of the stretched strings before they begin to dilute due to Hubble expansion is $m_s^2/H \gg m_s$, so corrections to \eqref{rhos} from string mode excitations are small.  See \appref{annulusapp} for details.}  The number density $n_c$ depends on the time derivatives of $z$ at the collision time, because the rate of change of the mass and the kinetic energy at the collision determines the amount of particle production.

We can incorporate the effects of string production and derive the homogeneous background equation for $z$ using the continuity equation $\dot \rho = -3H (\rho + p)$ as well as \eqref{rhos} and \eqref{effecac}:
\be \label{bkg1} \begin{split}
0 = 2 \gamma^3 \sigma \dot z \ddot z + 6 H \sigma \dot z^2 \gamma + V'(z) + {d \rho_s \over dt} + 3H \rho_s   \\ = 2 \gamma^3 \sigma \ddot z + 6 H \sigma \dot z \gamma + V'(z)+ f,
\end{split}
\ee
where $f(\dot z, \ddot z, ...) \equiv d\rho_s/dz$ is the ``force" due to string production, and the second line follows when $m_c \propto z-z_i$.

The force due to the background flux $V' \gg f$ in the early stages of inflation (although near the end of inflation when enough flux has been discharged this is no longer the case) so the term proportional to $f$ is a small correction to the background evolution.  Nevertheless we will retain it, as it plays an important role in the dynamics of the perturbations.

\paragraph{Perturbations:}
 The variation  in $f$ arises from several sources.  First, variations in the velocity $\dot z$ affect the amount of string production (we will neglect any dependence of $f$ on $\ddot z$ and higher derivatives).  Second, variations in the time of collision $\delta t_i = \delta z/\dot z_i$ affect the result.  Lastly, string or particle production is a quantum process, and there will be random variations $\delta n$ in the number density of produced strings.  Putting this together, the variation in $f$ is
\be \label{fpert} \begin{split}
\delta f = & \delta  \sum_{i} {\partial m_c \over \partial z} \, n_c \, e^{-3 H (t-t_{i})} \,  \Theta(t-t_{i}) = \sum_{i}e^{-3 H (t-t_{i})}  \left\{  \delta \dot z_i  \( {\partial^2 m_c \over \partial z \partial \dot z_i} \, n_c +{ \partial m_c \over \partial z} \, {\partial n_c \over \partial \dot z_i}\) \Theta(t-t_{i}) + \right. \\
& \left. \delta t_i {\partial m_c \over \partial z} \, n_c  \(3H \, \Theta(t-t_{i}) - \delta(t-t_{i}) \)  +  \delta n_c {\partial m_c \over \partial z} \,   \Theta(t-t_{i})\right\} \approx {\partial \bar f \over \partial \dot z}\,  \delta \dot z +m_0^2 \, \delta \bar n.
\end{split} \ee
Here,  $m_0^2 \equiv{ \partial m_c / \partial z}$, and the overbar and ``$\approx$" refer to a time average.  In what follows, we will solve the time-averaged equation, since the equation that results from keeping all the terms in \eqref{fpert} is difficult to deal with.  We will comment on the accuracy of this approximation shortly.

Using \eqref{fpert} and \eqref{bkg1}, after time-averaging the perturbed continuity equation is
\be \label{eq:pert1}
	\ddot{\del z} + 3 H(1 +  \lambda) \dot{\del z} -  e^{-2 H t} {\nabla^{2 } \over \gamma^2}   \del z
	= -{ m_0^2 \, \delta \bar n \over   2 \sigma \gamma^{3}} \, ,
\ee
where we have dropped a term proportional to $V''$, and $\lam$ is a dimensionless parameter controlling the strength of the friction due to the string or particle production in \eqref{eq:pert1}:
\be \label{lambda}
\lam \equiv \frac{\de_{\dot z}\bar f}{2 H  \sig \gam^3} \, .\ee

To proceed, we need to evaluate the source term $\delta \bar n$ that arises from fluctuations in the number density of produced particles or strings.  Assuming particle production is a Poisson process in physical space, the power spectrum of perturbations in the Fourier transformed number density  $\del n_{\k}$  is
\be \label{ns}
	\langle \del \bar n_{\k} \, \del \bar n_{\k'} \rangle =   {\avg{n} \over a^3} (2 \pi)^3 \del^{3}(\k+\k')  \, ,
\ee
where the factor of $a^{-3}$ is needed because $\k$ is comoving.  Eq. \eqref{ns} implies 
\be \label{deln}
\delta n_{\k}= \sqrt{\bar n} \, a^{-3/2} \, X_{\k} \, ,
\ee 
where $X_{\k}$ is a stochastic field satisfying $\langle X_{\k} \, X_{\k'} \rangle =    (2 \pi)^3 \del_D^{3}(\k+\k')$.

\paragraph{Power spectrum:}
Equation \eqref{eq:pert1} in momentum space describes a damped harmonic oscillator with an external source proportional to $\delta n$.
The general solution to \eqref{eq:pert1} is a linear combination of two independent solutions of the homogeneous equation plus a particular solution. The latter can be conveniently expressed in terms of a Green's function integral.
The homogeneous solutions (in terms of conformal time $\tau \equiv - e^{-H t}/H$) are
\be
\label{eq:solhom}
	\del z_{\k}(\tau) = C_{1, k} \tau^\nu J_\nu (c_s k \tau) + C_{2, k} \tau^\nu Y_\nu (c_s k \tau) \, ,
\ee
where $J_\nu$ and $Y_\nu$ are Bessel functions, and $\nu = (3+\lam)/2$.  

At late times $(\tau \to 0)$ the first term vanishes, so only the second term contributes to the power spectrum.
To fix the coefficients, we need to specify the initial conditions for the mode.
For frequencies higher than the energy of the produced particles, the particle production should not affect the modes of $\delta z$.    Therefore we can match each mode to the Bunch-Davies vacuum at a sufficiently large  value of the physical frequency $c_s k H \tau_e=M$ corresponding to the typical energy scale of the produced particles:
\be \label{delzk}
	\del z_{\k}(\tau_e) \sim \frac{i H}{\sqrt{4 \sig \gam^3 c_s k}} \tau_e e^{-i c_s k \tau_e} \, ,
\ee
which fixes
\be
	C_{2, k} = e^{- i \lam \pi/4} \frac{H \sqrt{\pi}}{2 \sqrt{2 \sig \gam^3}} \tau_e^{-\lam/2} \, .
\ee
The inhomogeneous contribution can be written as the integral of the Green's function
\be
	\del z_{\k} (\tau) = -   X_{\k} \int_{\tau_e}^\tau \rmd \tau' G(\tau, \tau') \frac{   m_0^2 \sqrt{\bar{n}}}{2 \sig \gam^3 \sqrt{- H \tau'}} \, ,
\ee
with
\be
	G(\tau, \tau') = \frac{\pi}{2} k c_s \tau \( \frac{\tau}{\tau'} \)^{\nu-1} \[ Y_\nu(k c_s \tau) J_\nu(k c_s \tau'))
	- J_\nu(k c_s \tau) Y_\nu(k c_s \tau' \] \, .
\ee
In the limit $\tau \to 0$ and $\tau_e \to - \infty$, the Green's function integral gives
\be
\label{eq:solinhom}
	\del z_{\k}(0) \to X_{\k} \frac{\pi}{\sqrt{H} (c_s k)^{3/2}} \frac{ m_0^2 \sqrt{\bar{n}}}{2   \sig \gam^3} \frac{\Gam(\nu)}{\Gam(\frac14) \Gam(\frac14+\nu)} \, .
\ee

The final power spectrum  therefore has two contributions, one proportional to the stochastic field $X_{\k}$ and the other due to the quantum fluctuations of the field.
These two contributions are uncorrelated, and so the power spectrum is the sum in quadrature of the two:
\be 
\label{eq:fullspectrum}
	P_\zeta(k) = \( { H \over M} \)^\lambda {2^{2 \nu} \Gamma(\nu)^2 \, H^4 \over 16 \pi^3 \sigma v^2} +\(\frac{\pi \Gam(\nu)}{\Gam(\frac{1}{4}) \Gam(\frac14+\nu)}\)^2
	\frac{ m_0^{4} \, \bar{n} H}{32 \pi^2 v^{2}  \sigma^2 \gamma^{3}}
	 \, .
\ee

When the friction due to string production is a significant effect ($\lambda \gg 1$),   the first term in \eqref{eq:fullspectrum} (arising from de Sitter fluctuations) is very small, and the power spectrum is dominated by the second term.  This was the regime considered in \cite{Green:2009ds,LopezNacir:2011kk}.In a string theory realization, $\lambda$ is generically very small for unwinding inflation, but string production may or may not be the dominant source of perturbations $\delta z$.

As a check, \eqref{eq:fullspectrum} is derived by solving \eqref{eq:pert1} directly for given $\delta n$ in \appref{ppapp}.  

\paragraph{Tensor power:} The tensor power has a contribution from de Sitter perturbations \eqref{tensor}.  As mentioned above, string or particle production can also contribute to ${\mathcal P}_h$ \cite{Senatore:2011sp}, but we leave this for future work.

\paragraph{Oscillations:}

Oscillations in the second term of \eqref{eq:fullspectrum}  arise due to the periodicity of the sum \eqref{rhos}.  We estimate the amplitude of these oscillations relative to the time-averaged result \eqref{eq:fullspectrum} in \appref{kickapp}.    We find that time averaging is a good approximation when $H \Delta t \ll 1$, where $\Delta t \equiv t_i - t_{i-1}$, because a high frequency driving force has very little effect on the mode.  Instead, if $H \Delta t \sim 1$, there are oscillations in the power spectrum due to string production with amplitude $\simleq 1$.  In the ``prototype" model where $\Delta t = l/2v \approx l/2$, the amplitude of the oscillations when $H l=1$ is $\approx 10^{-2}$, and then falls off as a high power of $H l$.

\paragraph{Tilt:} When the de Sitter fluctuations in \eqref{eq:fullspectrum} dominate the power spectrum, the tilt is given by \eqref{tilt}.  If instead the fluctuations from string production dominate, the tilt arises from the second term in \eqref{eq:fullspectrum}.  To evaluate it requires knowledge of $\nu, \bar n$, etc.  In string theory these are calculable, and the results can be found in \appref{annulusapp}.  The average density of strings $\bar n \sim g(b, v) m_s^{3}/H l$, where $g(b, v)$ is a dimensionless function of the velocity $v$ and impact parameter $b$, and $(H l)^{-1}$ is roughly the number of collisions per Hubble time.  When $b \gg m_s^{-1}$, the stretched open strings are never light, and string production is exponentially suppressed.  However, at least in the simplest cases one expects $b$ to be small or zero.  When $b=0$, $g \sim \O(10^{-1})$ (\figref{fplot}).

In general one expects a factor of $H^{-1}$ in $\bar n$, since the time-averaged density of produced particles is proportional to the number of collisions per Hubble time.  This means that the second term in \eqref{eq:fullspectrum} varies during inflation only through its dependence on $v$.  {So long as the dependence of $\bar n$ on $v$ is not too strong (as is the case in string theory for $0.5 < v <0.999$; see \figref{fplot}), its contribution to the tilt is small.  The prefactor involving $\Gamma(\nu)$ is very close to constant when $\lambda \ll 1$.

To evaluate the tilt due to the factor of $\gamma^{-3}$,
\eqref{bkg1} implies $\dot z \gamma \approx \gamma \approx V'/(6 \sigma H)$ when $\lambda \ll 1$.  At least in the ``prototype" model $H \propto Q$ and $V' \propto Q-1/2$ \eqref{eq:Vpp}, 
$$
 {\dot \gamma \over  \gamma} = {\dot Q \over Q(2 Q - 1)} \approx {\dot Q \over 2 Q^2}.
$$
Since $Q \approx Q_0 - 2 z/l$, $\dot Q = -2 \dot z/l$ and hence
$$
{d \ln \gamma^{-3} \over d \ln k} = - 3 {\dot \gamma \over H \gamma} = {3 \dot z \over H l Q^2} \sim {1 \over Q N},
$$
where we have used $N \sim (H l/ 2 \dot z) Q$ to derive the last expression.  Since $Q \gg 1$ the  tilt due to the factor $\gamma^{-3}$ is very small, and the spectrum of perturbations produced by strings is very close to scale invariant.

This conclusion relied on a number of assumptions, and could be modified depending on the microscopic details of the model.  For example, as mentioned above if $\sigma$ depends on the sizes of some compact dimensions and these change during inflation, there will be an additional contribution to the tilt.  Another possibility is that the dependence of $\lambda, m_0,$ and $\bar n$ on $v$ is not small and cannot be ignored.  

Still, at least in the examples we have investigated the tilt due to particle production is almost zero.  This has the interesting consequence that there will be an ``elbow" in the power spectrum:  if the red-tilted de Sitter fluctuations dominate in the early phase of inflation when CMB perturbations freeze out (as is probably necessary for consistency with the observational constraints on $n_s$), then at some shorter scale the string fluctuations will surpass them in amplitude, and from then on the tilt will be very close to zero.  Such a spectrum might have interesting observable consequences, for instance in so-called $\mu$-distortions~\cite{Pajer:2012dw}.

\subsection{Non-Gaussianity} \label{ngsec}

The effective theory described by \eqref{effecac} will be non-Gaussian because of the DBI kinetic term (the contribution from non-linearities in $V$ is  small \cite{Maldacena:2002vr}).  The shape and amplitude of the non-Gaussianity from a DBI kinetic term was described in \emph{e.g.} \cite{Senatore:2009gt}.  It is primarily equilateral:
$$
f_{\rm NL, equilateral} \sim (1-c_s^2)/c_s^2 \approx \gamma^2.
$$
The observational constraints on DBI inflation from WMAP require $c_{s} > 0.054$, or $\gamma < 19$. \cite{Senatore:2009gt}

If there are extra, approximately massless fields $\vec b$, the situation is more complex.  Assuming the fields $\vec b$ affect reheating -- as they do in the case of unwinding inflation, since they parametrize the transverse separation of the brane and anti-brane that must annihilate to end inflation -- they can reduce the amount of equilateral non-Gaussianity and increase the amount of local non-Gaussianity.  This is described in for instance \cite{RenauxPetel:2009sj} and \cite{Kidani:2012jp}.  Roughly speaking, the bigger the relative contribution of $\delta \vec b$ to $\zeta$, the less equilateral non-Gaussianity there will be (because $\vec b$ are non-relativistic), and the more local non-Gaussianity there will be.  
When $\langle \vec b \rangle = 0$ at the time of reheating, the $Z_2$ symmetry $\vec b \to -\vec b$ prevents $b$ fluctuations from affecting $\zeta$ at least to lowest order.   In this case neither the power spectrum not the bi-spectrum is affected by $\delta \vec b$, but higher correlations might be.  On the other hand if $\langle \vec b \rangle \neq 0 $, the $Z_2$ symmetry is broken and $\delta b$ is converted into $\zeta$ with some efficiency, therefore increasing the level of local non-Gaussianity and decreasing equilateral.

Another interesting feature that may arise in unwinding inflation is a peak at the ``folded" shape.  This occurs when there is particle production during inflation, or in general when inflation takes place away from the Bunch-Davies vacuum.  If particle or string production is a significant contribution to ${\mathcal P}_\zeta$, folded non-Gaussianity could be a smoking gun indication of it~\cite{pc}.

Oscillations in the power spectrum of the type predicted by unwinding inflation can lead to unexpectedly large ``resonant" non-Gaussianity~\cite{Flauger:2010ja}.
Finally, if $h(z)$ in \eqref{eq:effecach} changes rapidly (as is the case for unwinding inflation on $S_2$ -- see  \appref{S2app}), the non-adiabaticity may lead to additional non-Gaussianities.

\section{Stabilized compactifications} \label{casimir}

Unwinding inflation can occur in any model with at least one compact extra dimension threaded by a $(p+2) \geq 5$-form flux $F$, and containing a brane charged under $F$.   We have been assuming that the extra dimensions $\cal{M}$ are stabilized by some effect other than the flux that is discharged during the cascade, and that the initial energy density in the unwinding flux is small enough relative to the stabilization mechanism that the geometry of $\cal{M}$ is not strongly affected by the cascade.  

To show how this could happen, we give two examples.
In both, gradually decreasing the higher dimensional vacuum energy by discharging flux gradually decreases the 4D vacuum energy, as well as slightly decreasing the radius of the compact space. For a sufficiently slow decrease (combined with Hubble expansion to inflate away kinetic energy), the radion field corresponding to the radius of the compact space should remain very close to its local minimum.

\paragraph{$dS_4 \times S_1$:} 
 Stabilizing the $S_1$ can be accomplished with two ingredients:  a positive 5D cosmological constant $\Lambda_5$, and the Casimir energy of several bosons and fermions, at least some of which must be massive \cite{ArkaniHamed:2007gg}.  In our example, $\Lambda_5 = \Lambda_0 + \mu^5 Q^2/2$ is a combination of a ``bare" cosmological constant (which can arise from additional fluxes, vacuum energy, or other sources) and the flux that will discharge during the cascade.  

With these ingredients the effective potential for the radion of the $S_1$ can have a minimum, and the 4D vacuum energy $\Lambda_4(Q)$ in that minimum can be either positive, negative, or zero. 
If $\Lambda_4(Q_{0}) > 0$, the initial spacetime is metastable $dS_{4} \times S_{1}$.  At some time, a bubble of brane appears and begins to discharge $Q$.  For $Q_0 \gg 1$, this is a gradual process that slowly reduces $\Lambda_5$.   Because the minimum of the radion potential occurs due to a balance of negative Casimir energy against positive vacuum energy, reducing $Q$ has the effect of reducing $\Lambda_4$ and the radius of the circle.  Numerically, in the situation of interest the change in the radius of the circle is very small~(\figref{fig:casimir}) and can be neglected.  

\begin{figure}[ht]
\centering
	\includegraphics[width=0.45\textwidth]{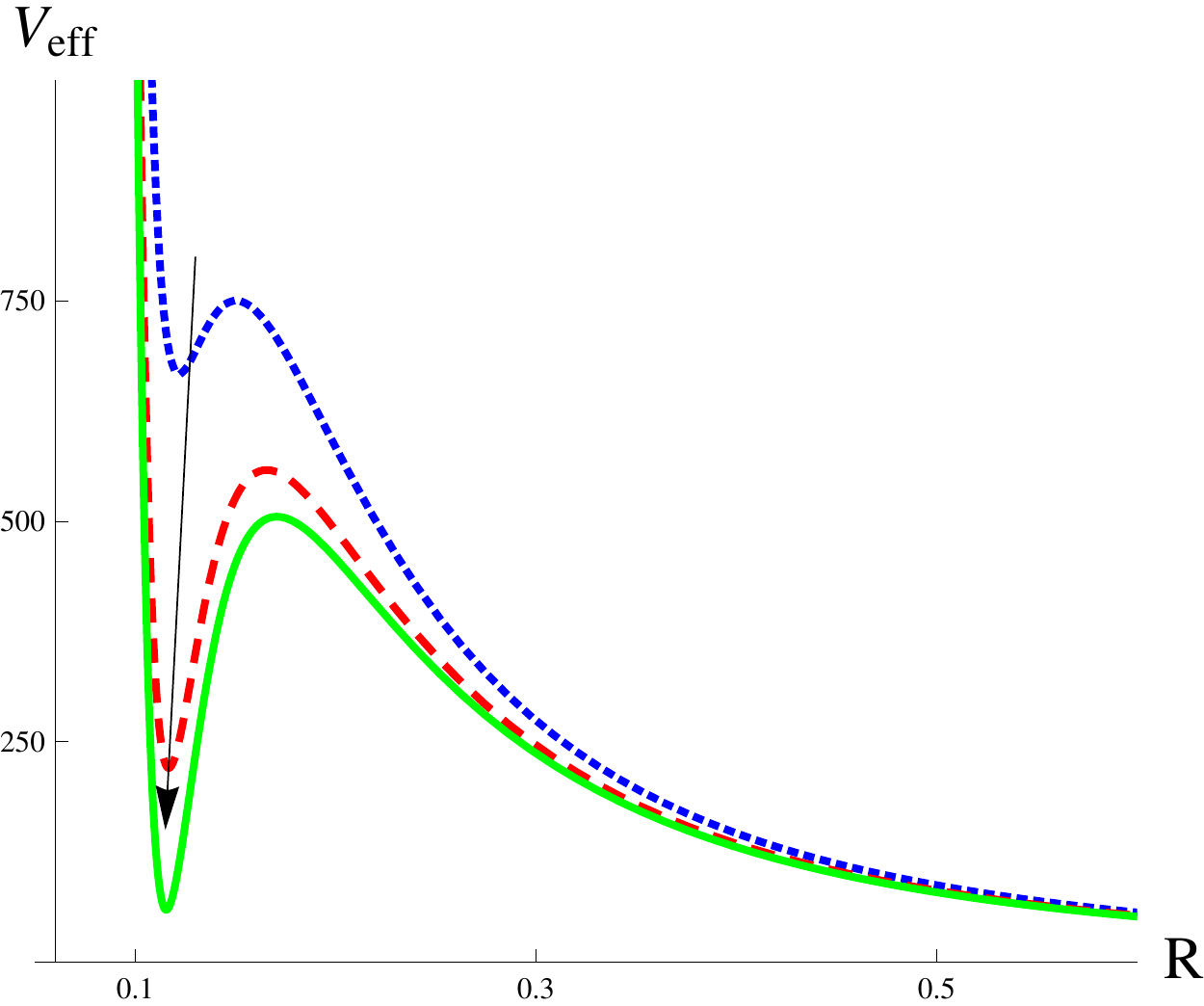}
	\includegraphics[width=0.45\textwidth]{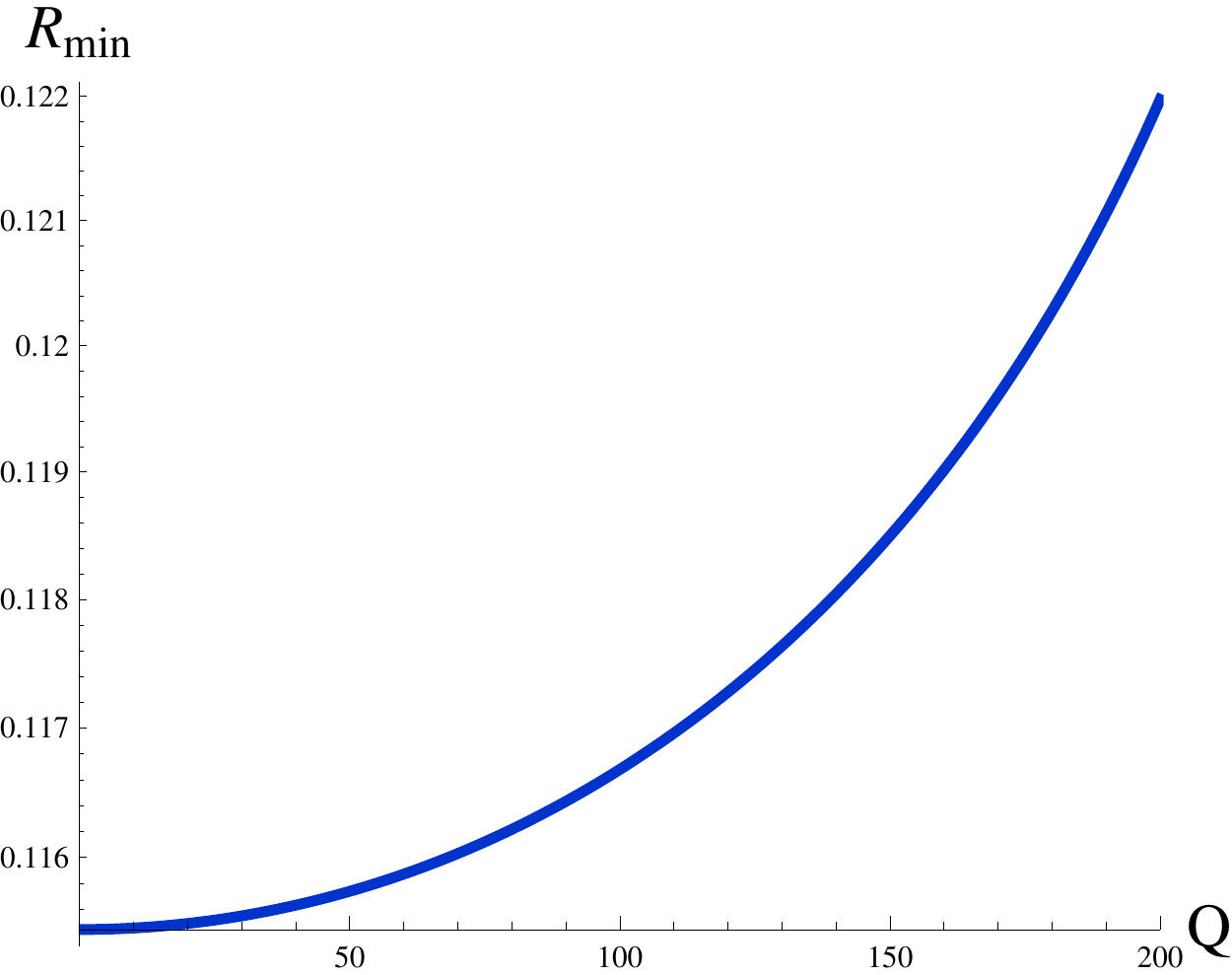}
	\caption{Stabilization of $S_1$. \textit{Left panel:} Effective potential for the radion field as the number of flux units decreases in the direction of the arrow. \textit{Right panel:} Change in the minimum of the radion as the number of flux units varies.}
\label{fig:casimir}
\end{figure}

The weak point of this model is that Casimir energy is a quantum effect, while the energy in flux is classical.  Hence this form of stabilization requires that quantum Casimir energy is larger than the energy in many units of classical flux.  This objection is alleviated to some extent by the fact that the extra dimensions in our model need only be slightly larger than the string length, so that quantum effects can be significant.  Furthermore in models with additional compact dimensions the flux quantization can be quite fine-grained, given only an $\O(1)$ hierarchy in their volume relative to the string or Planck length.

\paragraph{$dS_4 \times S_2$:} The $S_2$ can be stabilized by positive 6D vacuum energy and magnetic 2-form flux $F_2$ that threads the $S_2$~\cite{Freund:1980xh, Salam:1984cj}.   In this case the ``negative" energy term in the effective potential for the radius of the $S_2$ is provided by its curvature.  Once again, the 4D vacuum energy in the minimum can be positive, zero, or negative depending on the relative strengths of these contributions, and again, discharging part of the 6D vacuum energy by a flux cascade will gradually reduce the 4D vacuum energy.

\begin{figure}[ht]
\centering
	\includegraphics[width=0.45\textwidth]{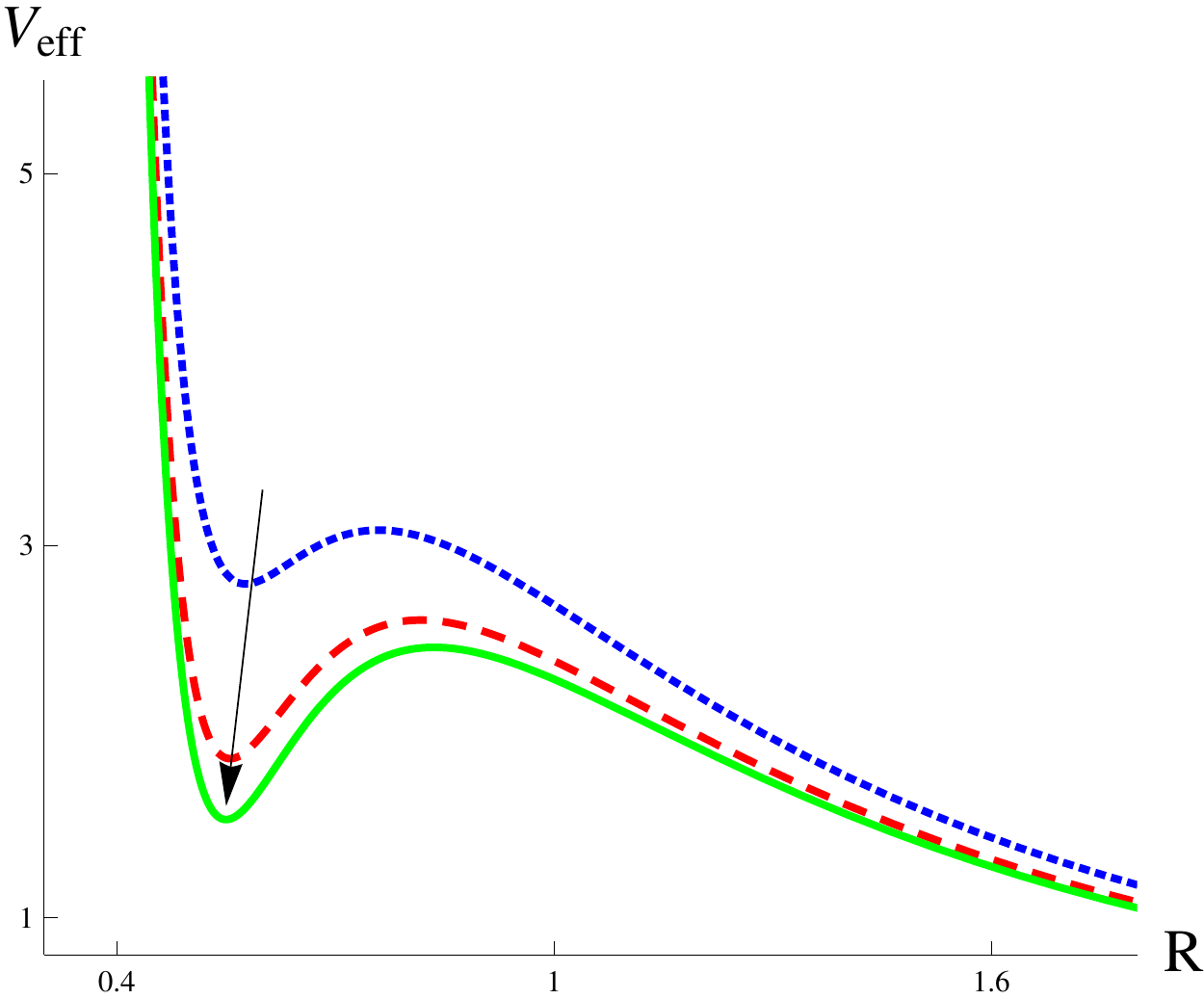}
	\includegraphics[width=0.45\textwidth]{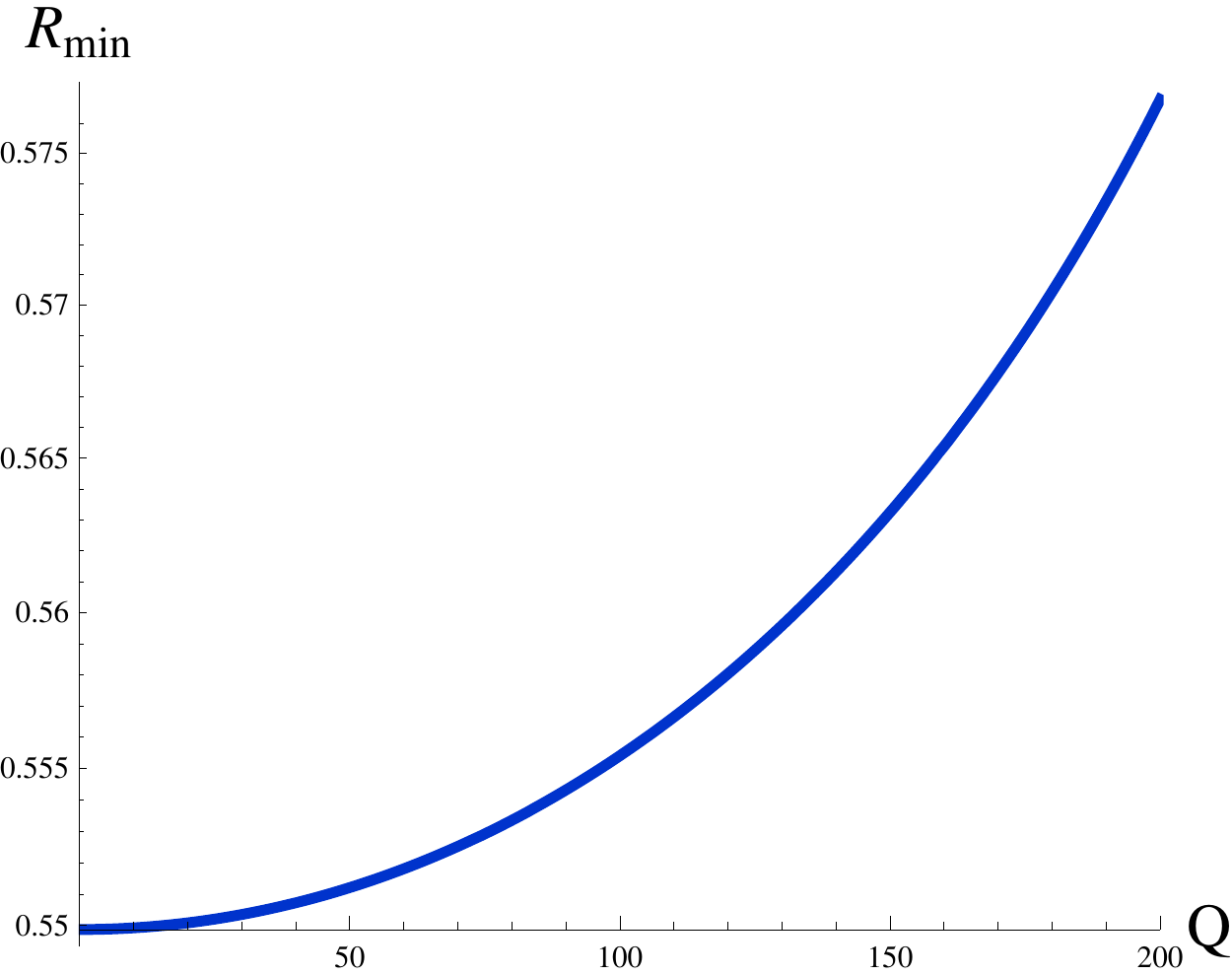}
	\caption{Stabilization of $S_2$. \textit{Left panel:} Effective potential for the radion field as the number of flux units decreases in the direction of the arrow. \textit{Right panel:} Change in the minimum of the radion as the number of flux units varies.}
\label{fig:flux}
\end{figure}

In this example, the flux cascade discharges a 6-form flux by the nucleation of a bubble of 4-brane, which is a circular string on the $S_2$~\cite{Kleban:2011cs}. The string oscillates back and forth from pole to pole on the $S_2$, wiping away an additional unit of 6-form flux with each pass.   More details can be found in \appref{S2app}.

\section{String theory} \label{stringsec}

The obvious context for unwinding inflation is string theory, which contains a variety of $D$-branes charged under Ramond-Ramond fluxes, as well as NS 5-branes charged under a 7-form flux.  However, realizing any model of inflation in string theory is extremely challenging, because the compact dimensions must remain at least approximately stable during 60 efolds of expansion of the non-compact dimensions.  Furthermore, supersymmetry is broken at least at scale $H$, making the theory difficult to control.
Here, we will only try to derive the  scalings of the parameters of our 4D effective description with $g_s$, $m_s$, and the geometry of the extra dimensions.  We will assume the compact geometry  remains stable throughout the cascade, and ignore any  couplings of the brane other than to the discharging flux. 

Unwinding inflation in string theory can in principle occur with a $p$-brane and flux $F_{p+2}$ with $3 \leq p \leq 8$.  As we will see however, it appears to work  most easily with $p = 4$ or $5$, and is probably not possible when $p=8$.

\subsection{Transverse volume}

In ten dimensions, there are $8-p$ transverse dimensions that are not threaded by the unwinding  flux.  Classically, the brane bubble does not expand or extend in these directions, because there is no flux forcing it to.  However, the relative position of the brane in the transverse directions (the impact parameter fields $\vec b$) has a very strong effect on string production during brane collisions (see \appref{annulusapp}), and in addition affects the time of  reheating and potentially produces interesting non-Gaussianity (see \secref{ngsec}).  

Furthermore, the size of the transverse dimensions determines the effective 4D brane charge.  When they are larger than the string scale, the energy per unit flux is reduced by ratios of the string scale to the size of the transverse dimensions \cite{Bousso:2000xa}.  
Because we require an initial flux number $Q_0 \simgeq \O(100)$, the characteristic length of the transverse extra dimensions must be somewhat larger than the string length in order to avoid super-stringy or super-Planckian energy densities---although as we will see, a ratio $\sim 5$ suffices for $p=5$.  The lack of transverse dimensions appears to rule out $p=8$, and makes $p=7$ problematic.  
Perhaps this conclusion can be modified by considering warping, but we will not do so here.

\subsection{The flux cycle}
The flux $F$ threads a non-contractible $(p-2)$ cycle in $\cal M$.  If the initial radius $R_0$ of the brane bubble when it nucleates  is small compared to the size of that cycle, the bubble will be roughly spherical and centered at a point on the cycle the flux threads.  After nucleating it will expand in all the flux directions and collide with itself as it wraps around,  initiating a flux cascade.  The geometry of the cycle is important in a number of ways, affecting both the average rate of discharge of background flux and the details of the perturbations (\appref{S2app}, \appref{torapp}).  There are too many such possibilities to discuss exhaustively here, so we will focus on some specific cases { (see \cite{Kleban:2011cs} for some additional examples).}

If the would-be radius of the bubble $R_0$ is larger than all the length-scales of the cycle, the dominant instanton will not have spherical topology or be localized at a point in the cycle.  Instead, it will be a sphere in the 3+1 non-compact dimensions of spacetime, but will wrap the cycle.  In this case there is no flux cascade---the bubble can be described by dimensional reduction, and so is just an ordinary 3+1D bubble inside of which the flux is reduced by only one unit.

If the cycle is anisotropic, the brane bubble may wrap a smaller sub-cycle but be localized and expand in other, larger directions \cite{Dubovsky:2011tu, Kachru:2002gs}.  For instance,  consider a cycle of the form $S_1 \times {\cal C}$.  If $l$ is the radius of the $S_1$ and $d$ is the characteristic length scale of $\cal C$, when $l>R_0>d$ the dominant instanton should wrap $\cal C$ and have spherical topology in the $dS_4 \times S_1$ directions.  In this case the dynamics can be described by dimensional reduction on $\cal C$, resulting in the ``prototype'' model of \secref{protosec} with an effective 3-brane with tension $\sigma \sim \sigma_p d^{p-3}$.  

Even if the cycle is not anisotropic and the initial bubble is small, it is possible that brane interactions could lead to an attractor solution where the brane wraps some subcycles.  For example, as the brane freely expands as a sphere on a torus $T_{q}$ it winds around more and more densely (its overall length increases as $z^{q-1}$).  Eventually, even at high velocity brane interactions may allow reconnection to a configuration with few or no self-intersections---namely a configuration where the brane wraps all but a single $S_{1}$ of the $T_{q}$.  If this quasi-equilibrium is reached before $60$ efolds from the end of inflation, the model will be effectively described as above.  A few  details of unwinding inflation on $T_q$ may be found in \appref{torapp}.

Another interesting case is that of $p=4$ where the 6-form flux extends in $dS_4 \times S_2$.  As described above, at least from the 6D point of view the $S_2$ can be stabilized by 2-form flux and a 6D vacuum energy, of which part is the 6-form flux $F$ that discharges during unwinding inflation.  The bubble of 4-brane is an oriented string on the $S_2$ that appears as a circle around some point.  The circular string expands, reaches its maximum extent at the equator of the $S_2$, and then contracts on the antipodal point from the one where it nucleated.  Assuming it does not annihilate, it will invert, and then again expand until it contracts on the point where it initially appeared.  Each such cycle discharges one additional unit of flux.  The 4D effective action for the brane and its perturbations is described in \appref{S2app}.

\subsection{Effective parameters from string theory}

In this subsection we will relate the parameters of our 4D description to those of the underlying string theory.  We will work in conventions where $m_s^2 = 1/(2 \pi \alpha') = 1/(2 \pi l_s^2)$.  
The fundamental tension and charge of a $p$-brane of type IIA/B string theory are 
$$
\sigma_p=\frac{m_s^{p+1}}{g_s(2\pi)^{(p-1)/2}}.
$$
When the $p$-brane wraps a $p-3$ cycle with volume $d^{p-3}$, the tension of the resulting effective 3-brane is 
$$
\sigma = \sigma_p d^{p-3} = \frac{m_s^{p+1}d^{p-3}}{g_s(2\pi)^{(p-1)/2}}.
$$
 
When all compact dimensions have radius $\sim d$, except an $S_1$ which has radius $l$, the 5D effective charge after reducing on everything but the $S_1$ is
$$
\mu^5 \sim \frac{\sigma M_{10}^{2p-14}d^{2(p-3)}}{d^5}=\frac{(d m_s)^{2p}g_s^{(p-3)/2}}{(2\pi)^2d^{11}m_s^6}.
$$
The Hubble constant during inflation is 
$$
H^2 \approx \ {8 \pi G_4 \over 3} {\mu^5 Q^2 l \over 2}.
$$

In order for the flux cascade to occur, we need  $R_0 \equiv 4 \sigma_{}/\kappa  \simleq l/2$, so that the dominant instanton is localized on the $S_1$ (as opposed to wrapping it).  For simplicity we will require $R_0 > d$ so that the brane wraps all but the $S_1$ of length $l$.  Furthermore we require that $Hd, Hl < 1$, on the prejudice that de Sitter compactifications with $d, l > 1/H$ are difficult or perhaps impossible to stabilize.  We require $l, d$ to be larger than the string length, and $N>60$ efolds of slow-roll inflation to solve the curvature problem. For consistency with observational constraints we require that the scalar density perturbations $P_\zeta$ have the observationally correct magnitude.

We can satisfy all these constraints simultaneously for $4 \leq p \leq 6$.  Qualitatively, $p=8$ is not possible because there are no transverse dimensions to dilute the energy per unit flux, and so one cannot have $Q_0 \gg 1$.  Similarly, the case $p=7$ is incompatible with these constraints because the single transverse dimension with length $d$ should satisfy $H d < 1$, while simultaneously being large enough to dilute the flux.  

{For $p=6$ the Lorentz factor is typically  large ($\gamma \sim 50$), and the parameters would require some tuning for consistency with the current constraints on non-gaussianity \cite{Senatore:2009gt}.}

For $p=3$ all the constraints above can be satisfied, but at the cost of a large hierarchy $l/d \sim 40$, and string production---which has a nearly flat spectrum---is the dominant source of perturbations.

\paragraph{Examples:} The two examples that work the most easily are $p=4$ and $p=5$. We give  a table (\tableref{parameters}) with a set of parameters for each case, and the resulting observational parameters. In both cases the scalar spectrum is dominated by de Sitter fluctuations rather than string production for $b\approx 0$: strings account respectively for $10\%$ of the power in the $p=4$ case and $1\%$ for $p=5$. Increasing the impact parameter $b$ suppresses the string contribution even more.

\begin{table}[t]
\centering
\begin{tabular} {l r}

\begin{tabular} {| l | l |}
	\multicolumn{2} { l } {$\bold{p=4}$} \\
	\hline
	$g_s = 0.01$ & ${\mathcal P}_{\zeta} = 2.4 \times 10^{-9}$ \\
	\hline
	$l = 20m_s^{-1}$ & ${\mathcal P}_{h} = 5.0 \times 10^{-11}$ \\
	\hline
	$d = 2.0m_s^{-1}$ & $r=2.1 \times 10^{-2}$ \\
	\hline
	$b\approx 0$ & $n_{s}-1 = -0.032$ \\
	\hline
	$Q_0=400$ & $H = 0.05 m_{s}$ \\
	\hline
	$Q_*=304$ & $\gamma = 11.5$\\
	\hline
\end{tabular}

\hspace{.6cm}

\begin{tabular} {| l | l |}
	\multicolumn{2} { l } {$\bold{p=5}$} \\
	\hline
	$g_s = 0.05$ & ${\mathcal P}_{\zeta} = 2.4 \times 10^{-9}$ \\
	\hline
	$l = 20m_s^{-1}$ & ${\mathcal P}_{h} = 1.4 \times 10^{-11}$ \\
	\hline
	$d = 4.9m_s^{-1}$ & $r=5.8 \times 10^{-2}$ \\
	\hline
	$b\approx 0$ & $n_{s}-1 = -0.032$ \\
	\hline
	$Q_0=400$ & $H = 0.04 m_{s}$ \\
	\hline
	$Q_*=314$ & $\gamma = 22.9$\\
	\hline
\end{tabular}

\end{tabular}

\caption{Parameter sets and the corresponding cosmological observables for $p=4$ and $p=5$ branes that expand around an $S_{1}$ with circumference $l$, and wrap one or two directions of a torus with circumference $d$.  Both sets satisfy all the constraints mentioned in the text.}
\label{parameters}
\end{table}

In both of the above cases, the oscillations in the energy density $V(z)$ are ${\mathcal O}(Q^{-2}) \sim 10^{-5}$, too small to be observed in the power spectrum.  However the oscillations in the perturbations due to string production are relatively large (see \appref{kickapp}).  If strings account for 10\% of the scalar power as in the example above with $p=4$, this could produce observable oscillations in the power spectrum.  Another interesting feature is the presence of an ``elbow'' in the power spectrum, where the perturbations due to strings overtake the red-tilted de Sitter perturbations at high $k$.


\section{Reheating}

When a brane and anti-brane pass within a string length of one another,  the lowest lying mode of the open strings stretched between them is tachyonic and will spontaneously condense.  If the relative velocity of the brane/anti-brane pair is relativistic, the branes will pass by each other and separate to more than a string length, in which case the lowest stretched string mode is no longer tachyonic (and one can meaningfully speak of its number density).  The calculation in \appref{annulusapp} shows that the density of such ``tachyons'' is  small in string units  when the velocity $v \simgeq .5$.  The conclusion is that the tachyon does not have time to condense when the velocity is relativistic; the same conclusion was reached by different methods in \cite{Kleban:2011cs}.

If instead the relative motion is slow and the brane and anti-brane spend a significant amount of time separated by less than a string length,  the tachyon can fully condense.  This is interpreted as the annihilation of the brane and anti-brane, with the $\sim 2 \sigma$ worth of energy density converted into closed string radiation.  Once the tachyon has condensed  the open string physics is strongly coupled \cite{Sen:1998sm}, and it is believed that the would-be open strings ending on the branes get confined into closed strings. \cite{Kleban:2000pf}

Other models of brane inflation use tachyon condensation as a mechanism for reheating, but unwinding inflation is unusual in that the brane/anti-brane pass close together many times before the tachyon condenses.  
At first, the flux is large enough to keep the branes moving relativistically.  Eventually, however, the flux decreases to the point that the branes move with non-relativistic velocity.  At some point after this, it will annihilate with an image anti-brane, reheating the universe and ending inflation.  This may occur at $Q=0$, or at some $Q_e$ satisfying $Q_0 \gg Q_e >0$.

A potential problem that could arise is if the brane sometimes over- or under-shoots $Q_e$ in some Hubble volumes, meaning that there are regions where the number of flux units remaining differs from  $Q_e$.  These regions will be surrounded by tensionful brane.  If they are sufficiently small, the tension of the brane will cause them to collapse and annihilate into radiation (or perhaps form black holes that subsequently evaporate).  But if they are sufficiently large and the energy density in them is lower than their surroundings, they may (depending on the tension, see \emph{e.g.} \cite{Chang:2007eq}) expand.  If so, they will collide and percolate, replacing the surrounding phase with the one inside them.  

This process will  create a feature in the power spectrum at the length scale corresponding to the typical separation between such ``overshoot'' regions.  Unless this length scale is $\sim e^{N_*} /H_e$ (where $H_e$ is the Hubble scale at the end of inflation) it is unlikely to affect CMB perturbations, but might have other interesting consequences.  Like percolating bubble collisions in first-order phase transitions, it could create gravity waves \cite{Kamionkowski:1993fg} or primordial black holes.  A detailed analysis of the model is required to determine this.

If the 4D vacuum energy corresponding to $Q=0$ is negative, then overshoot regions with sufficiently small or zero $Q$ may have negative vacuum energy.  If these regions expand, they may crunch the universe.
{Presumably, we live in the vacuum with the lowest $Q$ accessed by unwinding inflation.}

Models with very high re-heat temperature may suffer from a monopole problem, although the details of this are strongly model dependent.  Reheating may also produce cosmic strings, with potentially observable consequences.

\section{Conclusions}

From the ``bottom-up" or effective field theory point of view, inflation should be described by the simplest model consistent with data---perhaps single-field $m^2 \phi^2$.  On the other hand, natural phenomena sometimes turn out to be described by  field theories that are much more complex than initially seemed necessary, with the Standard Model  perhaps the prime example.
From the low-energy effective point of view,  unwinding inflation looks like an unholy union of $m^2 \phi^2$, DBI inflation, trapped/dissipative inflation,  hybrid inflation, and contains multiple light fields, an oscillating potential that may lead to resonant non-Gaussianity, and particle and graviton production.  But from the microscopic point of view it arises  from a very simple mechanism and seems to fit naturally into string theory, itself a theory based on one very simple assumption.  

Many issues remain largely unexplored. The dynamics of re-heating should be more carefully investigated.  The stability of the model---both of the compact extra dimensions and to other potential instabilities---needs to be studied.  If the formation of the initial bubble is not too rare,  collisions between two unwinding bubbles are possible, and their dynamics would be interesting to investigate.  The effects of the geometry of the compactification manifold on the power spectrum of perturbations have proven rich and interesting already at the superficial level we have investigated them.  One particularly important issue is the effect of the light scalars corresponding to the transverse directions on the fluctuations (both their amplitude and non-Gaussianity).  The spectrum of tensor modes and amount of dissipation  from brane Bremsstrahlung remains to be calculated.  Realizing unwinding inflation in string theory is a very important goal in its own right, and would help frame all of these questions.  

\section*{Acknowledgements}
It is a pleasure to thank A.~Brown, S.~Dubovsky, G.~Dvali, R.~Flauger, B.~Freivogel, V.~Gorbenko, A.~Hebecker, S.~Hellerman, S.~Kachru, N.~Kaloper, A.~Lawrence, J.~Maldacena, L.~McAllister, R.~Porto, M.~Roberts, L.~Senatore, T.~Tanaka, G.~Veneziano, T.~Weigand, and M.~Zaldarriaga for discussions.  M.K. and G.D'A. thank the 
G.D'A. is supported by a James Arthur Fellowship.
The work of MK is supported by NSF  grants PHY-1214302 and PHY-0645435.

\appendix

\section{Unwinding inflation on a sphere} \label{S2app}

In this section we consider a 4-brane on $dS_4 \times S_2$ with an initial six-form field strength $F_6 = \mu^{6/2} Q_0$ (where  $\mu$ is a  parameter with dimensions of mass, and $Q_0$ counts the units of flux).  The radius $R$ of the $S_2$ can be stabilized by 2-form flux; see \secref{casimir} for details.  We use the coordinates: 
\be
	ds^2 = -dt^2 +e^{2Ht} d\bold{x}^2 + R^2(d\theta^2 + \sin^2{\theta}d\phi^2)
\ee
and ignore the negative spatial curvature in the de Sitter directions.

The cascade will begin with the nucleation of a 4-brane extended in the 3 spatial de Sitter dimensions and co-dimension 1 on the sphere.  On the sphere, the brane will be an oriented, circular string with angular radius $\theta_b(t=0)$ centered on a random point that we choose to be $\theta=0$. When the bubble first forms it will contain $Q_0-1$ units of flux, but it will expand until it reaches the equator, contract on the point $\theta = \pi$, and then---assuming it doesn't annihilate---invert and expand, with opposite orientation, back towards the equator, contract to $\theta=0$, etc.  It will continue to oscillate between $\theta=0$ and $\theta=\pi$, discharging one unit of flux on each pass, until the cascade ends by annihilation.
 
The brane/flux action for this configuration is:
\be
\begin{split}\label{s2ac}
	S = &\int d^4x d\theta d\phi e^{3Ht}R^2\sin{\theta} \, \times \\
	&\( -\sigma\delta(R(\theta - \theta_b))\sqrt{1-R^2\( \dot{\theta_b}^2 - e^{-2 H t}(\nabla \theta_b)^2 - \frac{1}{R^2\sin{\theta}}(\partial_\phi \theta_b)^2 \)} - \frac{F_6^2}{2 \cdot 6!}\).
\end{split}
\ee

\paragraph{Background}
To solve for the background evolution, we treat the position of the brane as a function of time only: $\theta_b = \theta_b(t)$.  The action is:
\be
	S = \int d^4x d\theta d\phi e^{3Ht}R^2\sin{\theta} \( -\sigma\delta(R(\theta - \theta_b))\sqrt{1-R^2 \dot{\theta_b}^2 }- \frac{F_6^2}{2 \cdot 6!}\).
\ee

It is useful to introduce a monotonically increasing  variable $\omega_b$ to keep track of the number of passes that the brane has made over the sphere: $\omega_b = \theta_b$ for the first pass, $\omega_b=2\pi-\theta_b$ on the second, $\omega_{b}=2\pi + \theta_{b}$ on the third, and so on.  Using this variable, the energy density in the flux is:
\be
\begin{split}
	\frac{F_6^2(\theta)}{6!} =& \mu^6\left[\(Q_0-\[\frac{\omega_b}{\pi}\]\)^2\Theta\((-1)^{\[\omega_b/\pi\]}(\cos{\omega_b}-\cos{\theta})\)+ \right. \\
	&\left. \(Q_0-\[\frac{\omega_b}{\pi}\]-1\)^2\Theta\((-1)^{\[\omega_b/\pi\]}(-\cos{\omega_b}+\cos{\theta})\)\right] ,
\end{split}
\ee
where, again, $\[\dots\]$ denotes the integer part.

Performing the integrations over $\phi$ and $\theta$ 
and defining $z=R\omega_b$  gives
\be \label{s2action}
	S_4 = \int d^4x 2\pi e^{3Ht} \( -R\sigma |\sin{\frac{z}{R}}|\sqrt{1-\dot{z}^2} - V_4(z) \),
\ee
with 
\be \label{s2potential}
\begin{split}
	V_4 =& \frac{\mu^6R^2}{2}\(Q_0 - \[\frac{z}{\pi R}\]-1\)^2\(1-(-1)^{\[z/(\pi R)\]}\cos{\frac{z}{R}}\) + \\
	& \frac{\mu^6R^2}{2}\(Q_0 - \[\frac{z}{\pi R}\]\)^2\(1+(-1)^{\[z/(\pi R)\]}\cos{\frac{z}{R}}\) ,
\end{split}
\ee
(see \figref{fig:Vsphere}). 
The background equation of motion is
\be \label{spherebg}
	\ddot{z}+\frac{3H}{\gamma^2}\dot{z}+\frac{\cot{\frac{z}{R}}}{R\gamma^2}+\frac{V_4'(z)}{R\sigma \gamma^3|\sin{\frac{z}{R}}|}=0,
\ee
where $\gamma = (1-\dot{z}^2)^{-1/2}$ and $V_4'(z)=-\mu^6R|\sin{\frac{z}{R}}|(Q_0-1/2-\[\frac{z}{\pi R}\]) \equiv -\kappa|\sin{\frac{z}{R}}|$.

The force term proportional to $V'$ is the electric force, which pushes $z$ to increase until the flux is discharged and $Q_0-1/2-\[\frac{z}{\pi R}\]$ reaches zero or changes sign.  The curvature of the sphere induces an additional force term $\cot{(z/R)}/(R\gamma^2)$ that changes sign at the equator and has a possible divergence at the poles.  This arises from the fact that the brane's area is maximized at the equator and zero at the two poles.  

We have not been able to solve \eqref{spherebg} analytically, although it can be solved in terms of Jacobi functions when $H=V'=0$.  These  solutions (which describe a free string oscillating on the sphere without Hubble friction) are a good approximation near the inversion points $z = n \pi R$, and demonstrate that the kinetic energy $\gamma |\sin(z/R)|$ is finite and non-zero there, indicating that $\gamma \sim 1/|\sin(z/R)|$ near the inversion points.  Equation  \eqref{spherebg} can be solved numerically without difficulty (\figref{spheresol}).

\begin{figure}[b]
\begin{center}$
\begin{array}{cc}
	\includegraphics[width=0.5 \textwidth]{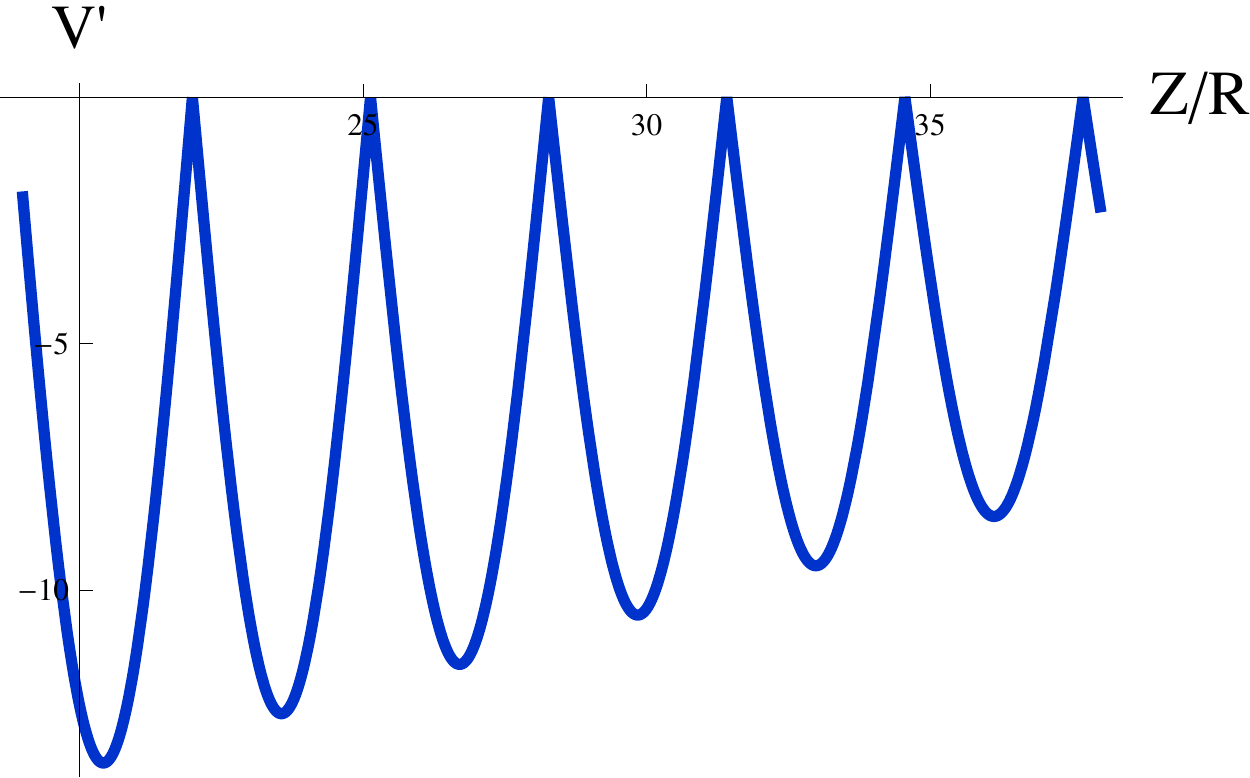} & \includegraphics[width=0.5 \textwidth]{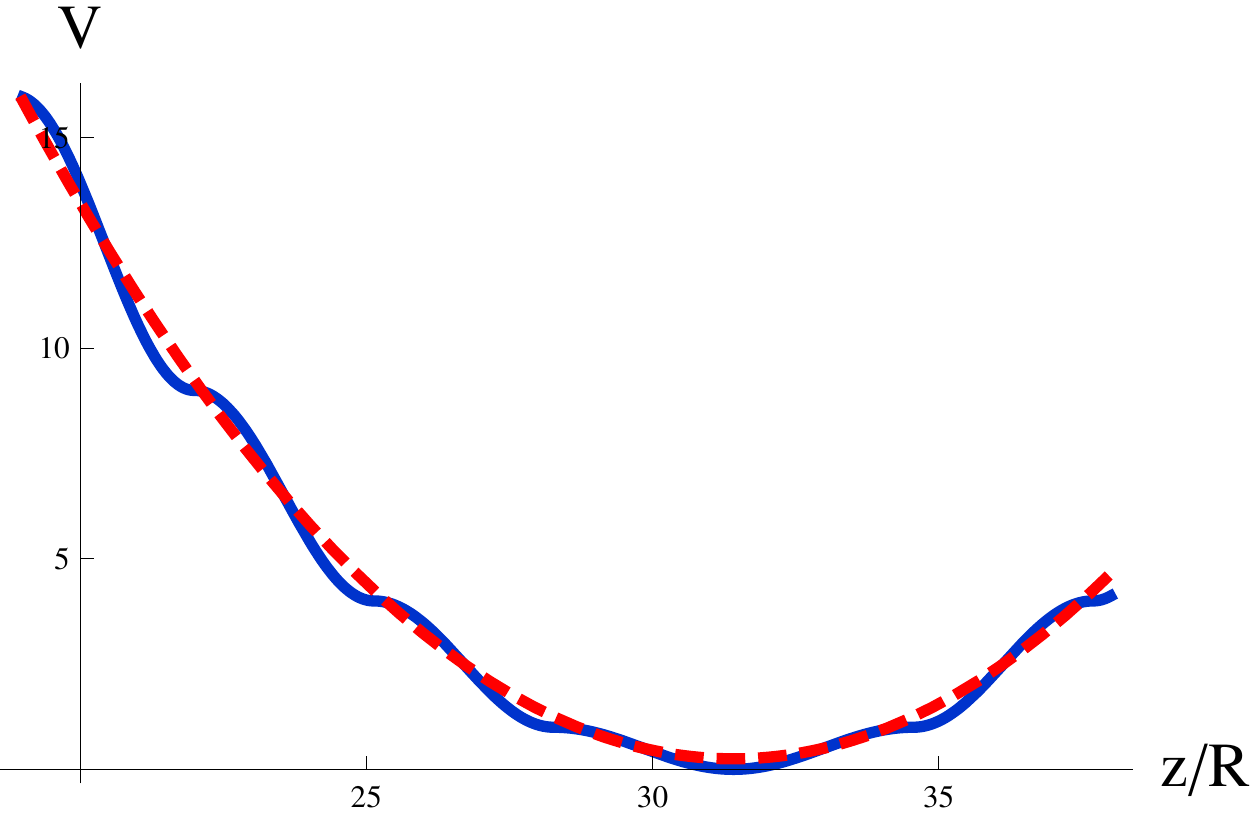}\\
\end{array}$
\end{center}
\caption{\label{fig:Vsphere}  Left panel:  The pressure $V'$ for the compactification on $S_2$.  Right panel:  The potential $V$ around the minimum; the smooth line is the quadratic approximation.}
\end{figure}

\begin{figure}
\begin{center}$
\begin{array}{ c c c}
	\includegraphics[width=0.32\textwidth]{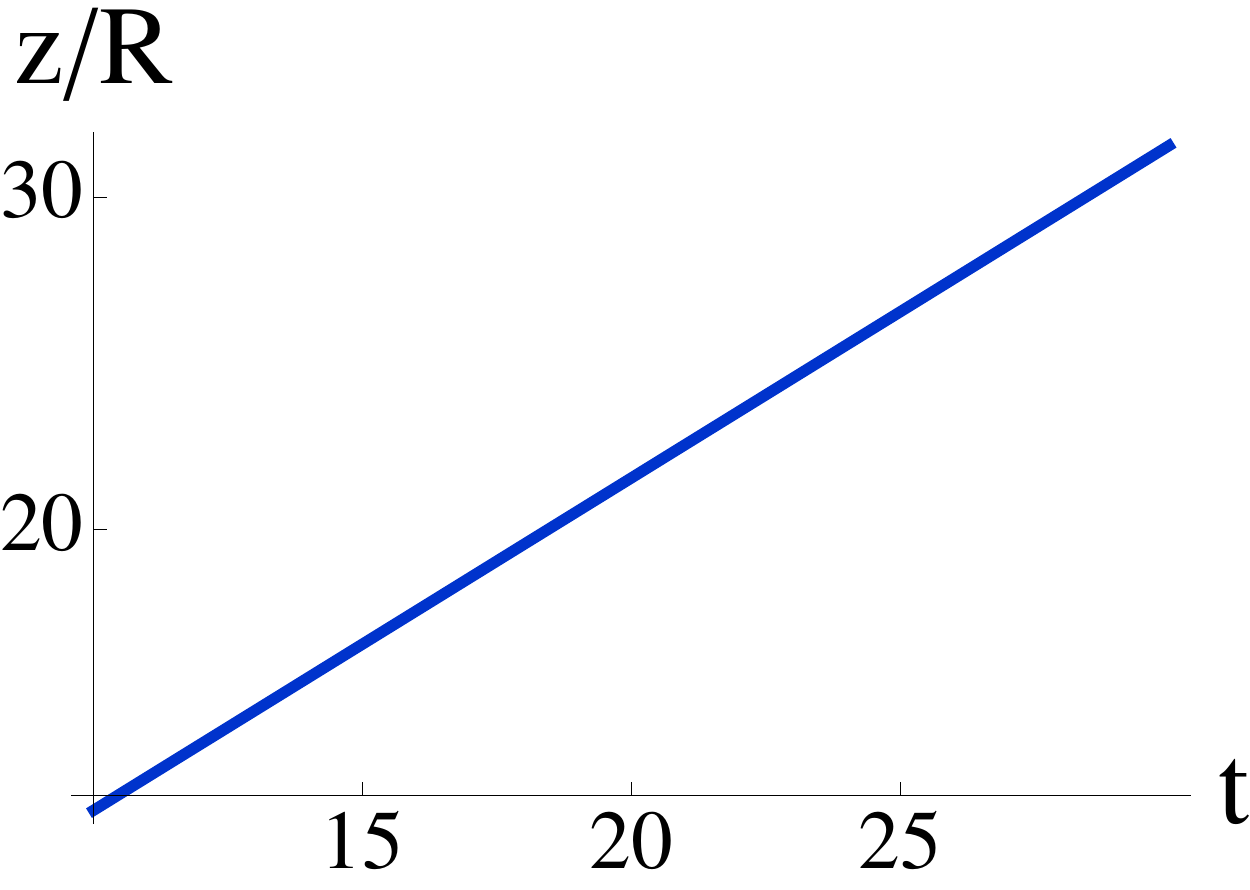} & \includegraphics[width=0.32\textwidth]{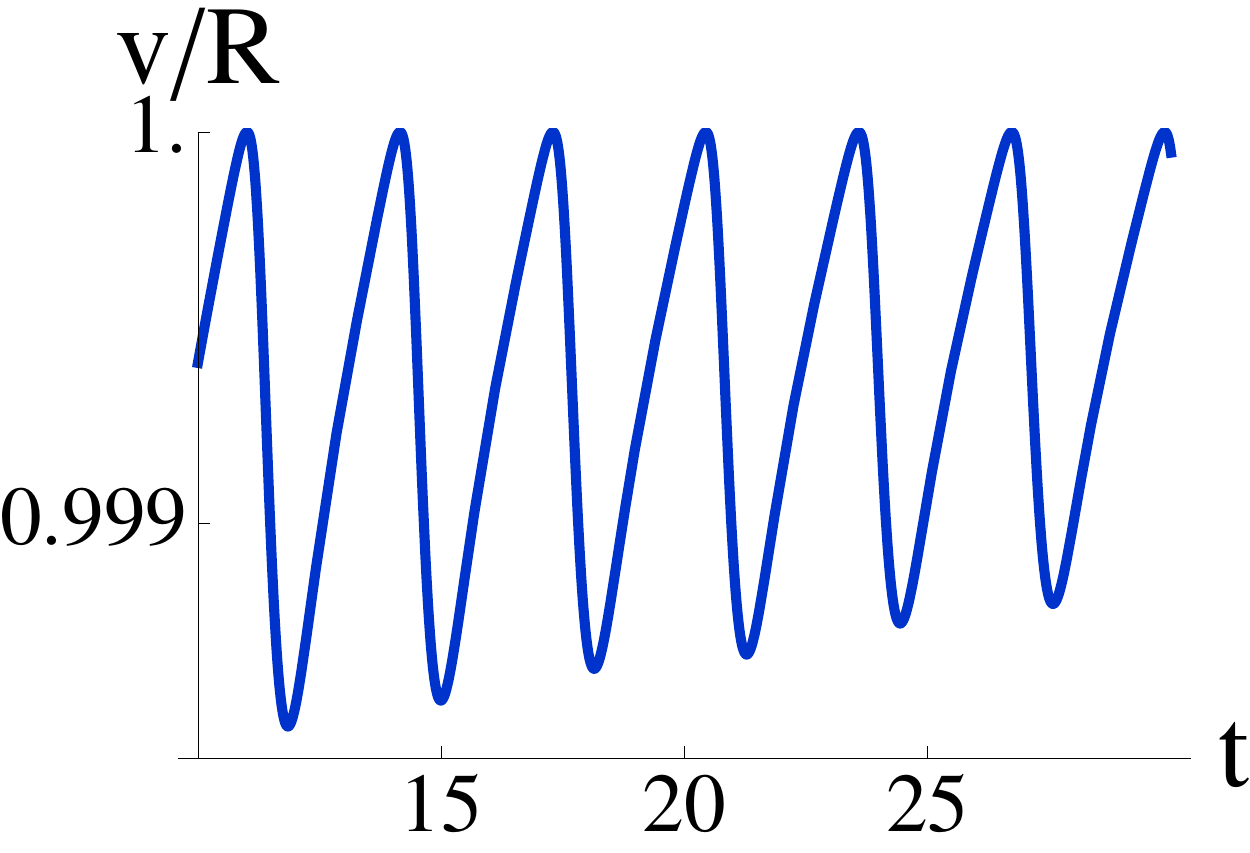} & \includegraphics[width=0.32\textwidth]{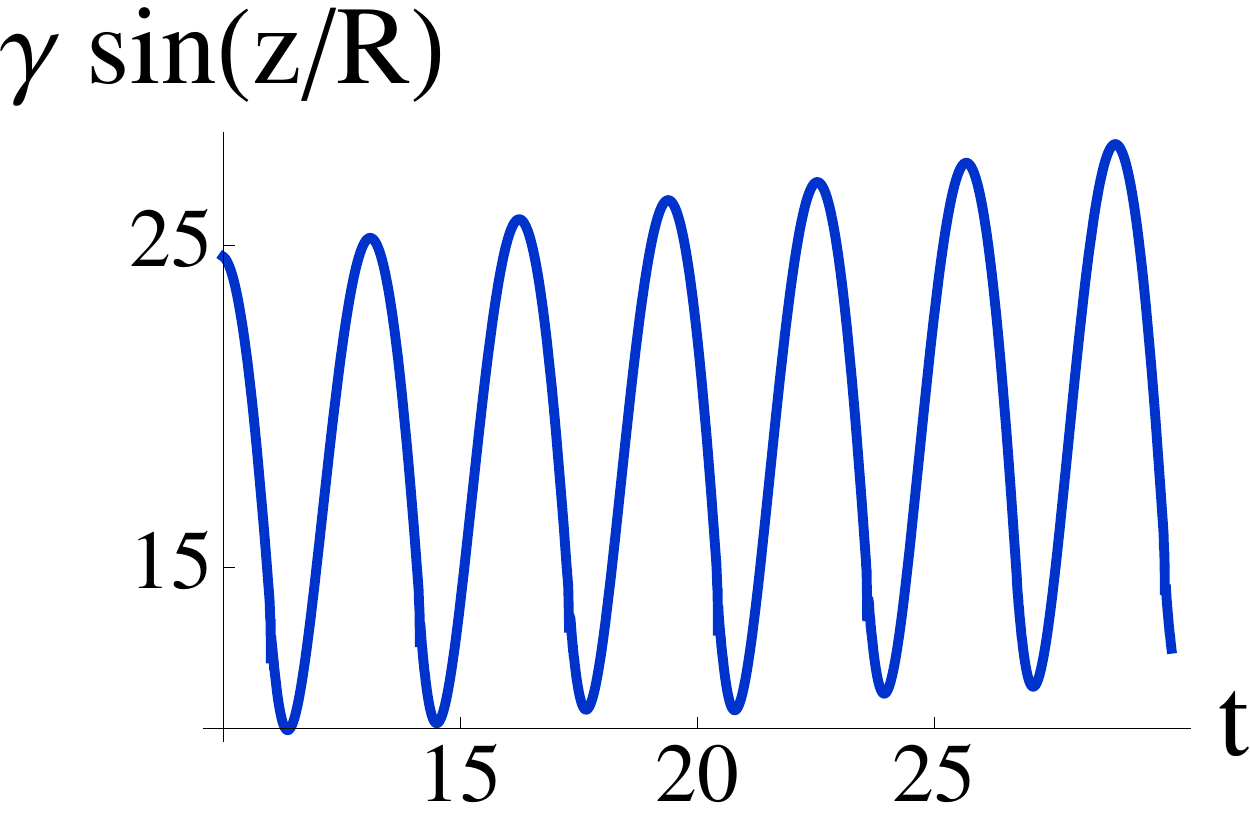} \\
\end{array}$
\end{center}
\caption{From left to right: the ``unwound'' angular position of the brane $\omega_{b}=z/R$, its angular velocity $\dot z/R$, and its Lorentz factor $\gamma$ multiplied by $|\sin(z/R)|$.}\label{spheresol}
\end{figure}

\paragraph{Perturbations}
To determine the perturbations around the smooth background evolution we define $\theta_b(t,\vec x ,\phi)=\theta_b(t)+\delta\theta_b(t, \vec x,\phi)$ and expand \eqref{s2ac} to second order in $\delta\theta_b$.  After performing the $\theta$ integration, the action for the perturbation is:
\be
\begin{split}
	S=& \int d^4x d\phi e^{3Ht} \[\frac{1}{2}R^3\sigma\sin{\theta_b}\gamma^3(\dot{\delta\theta_b})^2 + R^3 \sigma \cos{\theta_b}\gamma\dot{\theta_b} \delta\theta_b \dot{\delta\theta_b} \right.  \\ 
	& \left. \qquad  - \frac{1}{2}R^3\sigma\sin{\theta_b}\gamma\frac{(\partial_{\vec x}\delta\theta_b)^2}{e^{2Ht}} - \frac{1}{2}R\sigma\gamma(\partial_{\phi}\delta\theta_b)^2 + \( \frac{R\sigma\sin{\theta_b}}{2\gamma}-V''(\theta_b)\) \delta\theta_b^2 \].
\end{split}
\ee
  Expanding the $\phi$ dependence of the perturbation in modes:
$$
\delta z = \sum_n \delta z_n e^{i n \phi},
$$
where $z$ is defined as above.
Performing the $\phi$ integration gives:
\be
\begin{split}
	\int d^4x \pi  e^{3Ht} \left| \sin {z \over R} \right| &\[ R\sigma \dot{\delta z_n}^2 +2 \sigma\cot{\frac{z}{R}}\dot{z}\delta z_n\dot{\delta z_n} - R\sigma \gamma\frac{(\partial_{\vec x}\delta z_n)^2}{e^{2Ht}} \, +  \right. \\
	& \left. \(\frac{\sigma }{R\gamma}-\frac{\sigma\gamma n^2}{R \left| \sin{z \over R }\right|}-{V''(z) \over  \left| \sin{z \over R }\right|}\)\delta z_n^2 \],
\end{split}
\ee
Dropping the subscript, the  equation of motion is:
\be \label{spert}
\begin{split}
&\ddot{\delta z} + \[3H +\dot{z}\(\frac{\cot\frac{z}{R}}{R}+3\gamma^2\ddot{z}\) \] \dot{\delta z}\, +\\
& \quad  \[\frac{-1}{R^2\sin^2\frac{z}{R}\gamma^2}+\frac{k^2}{\gamma^2e^{2Ht}}+\frac{n^2}{R^2\gamma^2 \left| \sin{\frac{z}{R}} \right| }+\frac{1}{R\sigma\gamma^3}\frac{\p}{\p z}\(\frac{V'(z)}{\left| \sin\frac{z}{R} \right| }\)\]\delta z=0.
\end{split}
\ee
This equation is somewhat difficult to deal with numerically, primarily due to the coefficient of $\delta \dot z$.  However, at superhorizon wavelengths ($k=0$),  $\delta z = \dot z$ is an exact solution to \eqref{spert}.  Because the time-averaged friction (the time average of the coefficient of $\delta \dot z$) is positive, this should be the ``growing'' solution, a conclusion  we have verified numerically.  Since $\dot z$ is a constant plus small oscillations, the perturbations are nearly constant outside the horizon.  To determine their amplitude requires numerically solving \eqref{spert} for finite $k$.  Preliminary  investigations  indicate that their average amplitude is parametrically similar to the $S_{1}$ case \eqref{eq:noSspectrum}, but with additional oscillations with amplitude depending on $H R$.

Calculating the degree of string production when the brane turns inside out is an interesting problem that we leave for the future.  We tentatively expect that the results will again not be substantially different from that of parallel brane/anti-brane scattering, but this remains to be established.

\section{Unwinding inflation on a torus} \label{torapp}

Here we consider the case of a flat $q$-torus, where the bubble nucleates with a diameter smaller than any cycle of the torus.  Ignoring  interactions for now,  the brane will expand freely and uniformly in all $q$ directions with velocity $v \sim 1$, wrapping repeatedly around the cycles of the torus and intersecting itself.  Its overall area will increase as $\Omega_{q-1} z^{q-1}$, the area of a $q-1$ sphere of radius $z$.

The flux will discharge in a complicated pattern of overlapping spheres, but the average number of flux units discharged is simple to calculate.  In the covering lattice of the torus, when the radius is $z$ the wall of any image bubble that nucleated within a distance $z$ of the origin will have crossed the origin and discharged a unit of flux.  Since there are $\approx (\Omega_{q-1}/q) (z/l)^{q}$ such points, the flux 
$$
Q(z, \vec y) = Q_{0} - {\Omega_{q-1} \over q} \({z \over l} \)^{q}\(1 + r(z, \vec y)\),
$$
where $\vec y$ are the coordinates on the $T_{q}$, $l^{q}$ is the volume of the torus, and  $r(z, \vec y)$ is the error in approximating the number of points within a sphere by the volume of the sphere over the volume of the torus. The function $r(l, \vec y) \sim \O(1)$, it oscillates and falls off like a negative power of $z/l$ for $z \gg l$ and it averages to zero over $\vec y$ :  $\int d^{q}y \, r(z, \vec y) = 0$.

Dropping the small oscillating remainder term $r$, the action is
\be
S \approx -\int dt d\vec x e^{3 H t} \left\{  \sigma_{q+2} \Omega_{q-1} z^{q-1} \sqrt{1-\dot z^{2}}  + \mu^{4+q} l^{q} \(Q_{0} - {\Omega_{q-1} \over q} \({z \over l} \)^{q}\)^{2} \right\}.
\ee
The equation of motion is
$$
\ddot z + {3 H \dot z \over \gamma^{2}} + {q-1 \over \gamma z} + {V' \over \gamma^{3} \Omega_{q-1} \sigma z^{q-1}}=0.
$$
The power spectrum of de Sitter fluctuations is \eqref{eq:noSspectrum}
$$
{\mathcal P}_{\zeta} \approx {H^{4} \over 4 \pi^{2} \Omega_{q-1} \sigma_{q+2} v^{2} z^{q-1} }.
$$
Physically, the $\sigma z^{q-1}$ in the denominator represents the fact that the effective mass of the brane is increasing as it expands on the torus.  The larger the mass, the greater the inertia and the smaller the amplitude of  de Sitter fluctuations.

The tilt is
$$
n_{s}-1 \approx -{(q-1) v \over H z} + 4 \frac{\dot H}{H^2},
$$
which is red and relatively large in magnitude.

\section{String production from brane/anti-brane scattering} \label{annulusapp}

In this appendix we  derive the amount of open string production between a parallel $p$-brane and  anti-$p$-brane that pass each other with impact parameter $b$ and velocity $v$.\footnote{A calculation using effective field theory in a related scenario appears in \cite{Brandenberger:2008kn}.}  After a few efolds of unwinding inflation on $S_1$ the radius of the brane bubble is exponentially large, so that parallel branes are a good approximation (see \figref{prl}).  We perform the calculation in flat space, ignoring the de Sitter curvature---but since string production takes place in a period $l_{s}/v \ll H^{-1}$, we expect this is also a good approximation.

We ignore closed string production, which may occur via brane Bremsstrahlung or by decay of the open strings that are produced.    At ultrarelativistic velocities and large brane co-dimension, closed string radiation is a very important effect\footnote{We thank L. McAllister for discussions on this point.} and neglecting it may not be justified.  However, Bremsstrahlung is proportional to acceleration, and therefore should not affect the motion of the brane at its terminal velocity \eqref{velocity}.  In principle stretched branes could also be produced, but  the rate should be suppressed relative to open string production

At lowest order the rate of string production in brane scattering is determined by the imaginary part of the annulus diagram with the appropriate boundary conditions \cite{Bachas:1995kx}.  Physically, string production occurs because the mass of the stretched strings has a contribution that is proportional to length, which is changing with time as the branes move.  The calculation is T-dual to the pair production of open strings in an electric field, \cite{Bachas:1992bh} and has many features in common with the Schwinger effect.

Our analysis differs from \cite{Bachas:1995kx} in that we consider brane/anti-brane scattering, rather than brane/brane.  This has important consequences at low velocity:  the brane/brane case becomes supersymmetric as $v \to 0$, and the rate of string production scales as $v$ to a positive power.  This is related to the fact that the lightest stretched open string mode is massless when the branes coincide.  For brane/anti-brane  scattering, there is a tachyonic  mode that will condense if the brane and anti-brane remain in close proximity for long enough.  As we will see, relativistic velocity prevents this from happening even when the impact parameter $b=0$:  the lightest mode is tachyonic only briefly, and does not condense.  Instead, a finite density of strings are created in this mode.

Our starting point is equation (11) of \cite{Bachas:1995kx} for the phase shift for forward scattering:
\be \label{delta}
	\delta(b,v) = -2\frac{V_p}{4\pi}\int_0^{\infty}\frac{dt}{t}(2\pi^2)^{-\frac{p}{2}}e^{-\frac{m_s^2b^2t}{2}}\frac{\theta_1'(0 | \frac{it}{2})}{\theta_1(\frac{\chi t}{2\pi} | \frac{it}{2})}\sum_{\alpha=2,3,4}\frac{1}{2}e_{\alpha}\theta_{\alpha}\(\frac{\chi t}{2\pi} | \frac{it}{2}\)\theta_{\alpha}^3\(0 | \frac{it}{2}\)\eta^{-12}\(\frac{it}{2}\)
\ee
where $V_p$ is the volume of the $p$-brane (or anti-brane), $\theta$'s and $\eta$ are the Jacobi and Dedekind functions for the definitions and properties of which we address the reader to~\cite{Polchinski:1998rq}, $\chi = \tanh^{-1}v = \tanh^{-1}v_{2}-\tanh^{-1}v_{1}$ is the relative rapidity, and $b = | \vec b|$ is the impact parameter. The charges $(e_{\alpha})$ in the brane-anti brane case are such that $e_2=-e_3=-e_4=-1$.

 We are interested in the imaginary part of \eqref{delta}. The poles of the integrand are the zeros of $\theta_1(\chi t/2\pi | it/2)$, \emph{i.e.}~$\chi t/2\pi = k \in \mathds{Z}_0$.  The residue of the $k$-th pole is $\pi (-1)^k$. Therefore:
 \be
 	\textrm{Im}(\delta) = -\frac{i}{2}\frac{m_s^pV_p}{(2\pi)^p}\sum_k(-1)^k\frac{\chi^{p/2}}{k^{p/2+1}}e^{-\frac{\pi m_s^2b^2k}{\chi}}\sum_{\alpha=2,3,4}\frac{1}{2}e_{\alpha}\theta_{\alpha}\(k | \frac{i\pi k}{\chi}\)\theta_{\alpha}^3\(0 | \frac{i\pi k}{\chi}\)\eta^{-12}\(\frac{i\pi k}{\chi}\) ,
 \ee
 which, using properties of the Jacobi functions, can be rewritten as
\be
\label{ampl}
\begin{split}
	\textrm{Im}(\delta) =  \frac{-m_s^pV_p}{2(2\pi)^p}\sum_k\frac{\chi^{p/2}}{k^{p/2+1}} e^{-\frac{\pi m_s^2b^2k}{\chi}} & \bigg\{ {\eta^{-12}\(\frac{i \pi k}{\chi}\)}\bigg[(-1+(-1)^k)\theta_3^4\(0 | \frac{i\pi k}{\chi}\) \\
	 & +(1+(-1)^k)\theta_4^4\(0 | \frac{i \pi k}{\chi}\)\bigg] \bigg\} .
	 \end{split}
\ee

We want to expand the term in the curly brackets around small $q \equiv \exp(2\pi i (i\pi k/\chi))$ to reveal the string mode expansion.
One has
\be \label{expansion}
	{\eta}^{-12}(q)\[(-1+(-1)^k){\theta}_3^4\(q\)+(1+(-1)^k){\theta}_4^4\(q\)\]  = \frac{2(-1)^k}{\sqrt{q}}-16+72(-1)^k\sqrt{q}-256q+O(q^{3/2}) .
\ee
The integer coefficients are the number of stretched open string degrees of freedom at the corresponding oscillator level---a complex scalar tachyon, 16 massless fermionic degrees of freedom, etc.---while the factors of $(-1)^k$ arise due to spin statistics as usual for the Schwinger effect (\emph{c.f.} \cite{Bachas:1992bh}).

Ordinarily the sum over $k$ is almost irrelevant, because the higher terms are exponentially suppressed.  In our case the sum looks problematic due to the term that corresponds to the tachyon (at least for small $b$ where the string is actually tachyonic at closest approach):
$$
\sum_{k=1}^{\infty} (-1)^k k^{-p/2-1} \exp\left[{\pi k \over \chi}\(\pi  -m_s^2 b^2 \) \right].
$$
However, even for $b=0$  the sum can be performed exactly in terms of a polylogarithm, and (with the help of the $(-1)^k$ factor) is real and finite  away from $v=\chi=0$.  The physical interpretation is that when $v \sim 1$ the stretched string is only very briefly tachyonic and does not have time to condense.  By contrast for $v \ll 1$ there is a divergence, which  leads to brane-antibrane annihilation due to tachyon condensation.

From \eqref{ampl}, the amplitude given by the  four lightest open string modes (tachyonic, massless and the first two massive) is:
\be
\begin{split} \label{amplfin}
	\textrm{Im}(\delta) \simeq  & -\frac{1}{2}\frac{m_s^pV_p}{(2\pi)^p}\chi^{p/2}  \left[-2\textrm{Li}_{\frac{p}{2}+1}(-e^{\frac{-\pi m_s^2b^2+\pi^2}{\chi}}) +16\textrm{Li}_{\frac{p}{2}+1}(e^{-\frac{\pi m_s^2b^2}{\chi}}) \right. \\
	& \left.-36\textrm{Li}_{\frac{p}{2}+1}(-e^{-\frac{\pi m_s^2b^2+\pi^2}{\chi}}) + 256\textrm{Li}_{\frac{p}{2}+1}(e^{-\frac{\pi m_s^2b^2+2\pi^2}{\chi}}) + \O\(e^{-3 {\pi^2 \over \chi}} \)\right] \\
	& \equiv -\frac{1}{2}\frac{m_s^pV_p}{(2\pi)^p}\chi^{p/2} F(b,\chi) .
\end{split}
\ee

The  energy density in the produced strings is simply related to \eqref{amplfin}: we should multiply by the energy per string and divide by the volume.  The energy per string is evidently \cite{McAllister:2004gd}
\be
	E^2 = \frac{v^2}{\chi^2} \[m_s^4(z^2+b^2) +  \pi m_s^2 (n-1) \] \approx  \frac{v^2}{\chi^2} m_s^4(z^2+b^2),
\ee
where $\sqrt{z^2+b^2}$ is the length of the string and the integer $n$ is the oscillator number (in our notation $n=0$ is the lowest mode of the stretched string, which is tachyonic when $z=b=0$).  
The pre-factor $v/\chi$ can be derived by matching to a field theory computation  \cite{McAllister:2004gd}.  In any case (because $\chi \sim \ln \gamma$) it is not an important effect except for ultra-relativistic scattering, and reduces to the standard form when $v \ll 1$. 
The approximation in the last step holds for unwinding inflation, where the typical value of $z \gg m_s^{-1}$ and only the first few oscillator modes are produced.

The energy density of  strings produced at a single collision is then
\be
	\rho_{c}(\chi,z) = E \, {\text Im}(\delta)/V_p  \approx \frac{1}{2}\frac{m_s^{p+2}}{(2\pi)^p}\chi^{p/2}\frac{v}{\chi}F(b,\chi)\sqrt{z^2+b^2}.
\ee
For $0.9 < v < 0.999$ and $b \approx 0$ the quantity $\frac{1}{2}\frac{1}{(2\pi)^p}\chi^{p/2}\frac{v}{\chi}F \sim (2 \pi)^{3-p}$, as is its derivative with respect to $v$ (\figref{fplot}).  For $g_s \ll 1$, this means that the energy density in produced strings is much less than the brane tension $\sigma \sim m_s^{p+1}/g_s$ at least for  $z \sim l_s$.  For $z \gg l_s$ the energy density in stretched strings will exceed the brane tension, but during unwinding inflation the brane continues to move at nearly constant $v$ due to Hubble dilution of the strings and the force from the background flux; see \secref{ppsec}.

\begin{figure}[t]
\begin{center}
	\includegraphics[width=.6\textwidth]{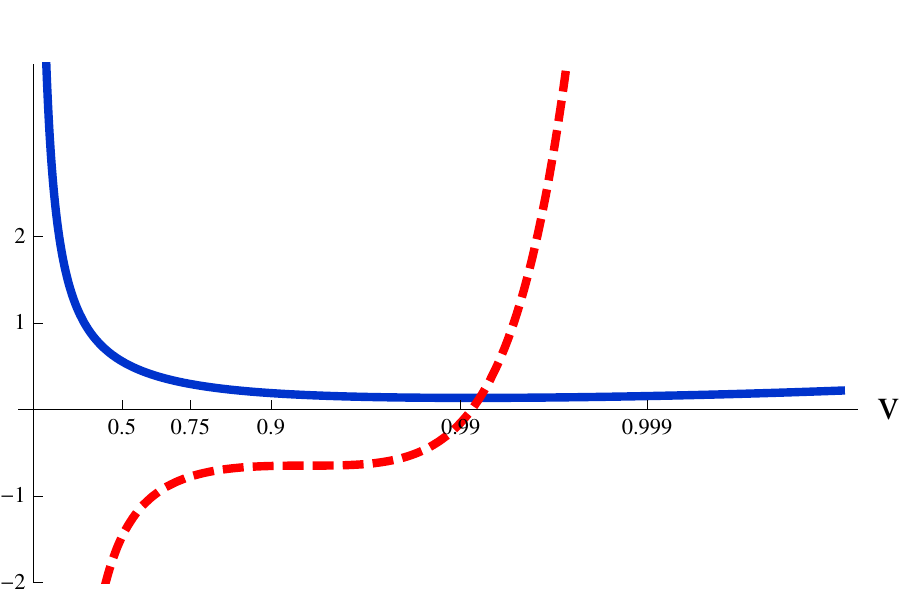}
\end{center}
\caption{The quantity $n_cm_0^2/(2\pi^{3-p})$ (solid line) and its derivative with respect to $v$ (dashed line),  in units  $m_s=(2 \pi l_{s}^{2})^{-1/2}=1$.  Both  are $\O(1)$, with only a mild dependence on $v$  for $0.5<v<0.999$.  The case shown is $p=4$, but the magnitude and shape are very similar for other $p$.  The divergence as $v \to 0$ is due to the tachyon; the divergence as $v \to 1$ is due to the copious production of strings at ultrarelativistic velocities. }\label{fplot}
\end{figure}

\noindent In the notation used in \secref{ppsec}:
\begin{eqnarray}
	&m_0^2 = \frac{\p m_c}{\p z} = m_s^2\frac{v}{\chi} \,\,\,\,\,\,\,\,\,\,\,\,\,\,\,\,\,\,\,\,\,\,\,\,\,\,\,\,\,\,\,\,\, & n_c= \frac{m_s^p}{2(2\pi)^p}\chi^{p/2}F(b,\chi) \nonumber \\
	&\lambda = \frac{\p_{v}\bar{f}}{2H\sigma\gamma^3}	 \,\,\,\,\,\,\,\,\,\,\,\,\,\,\,\,\, \,\,\,\,\,\,\,\, \,\,\,\,\,\,\,\, & \bar{f} = \frac{1}{3H\Delta t}\(\frac{\p m_c}{\p z}n_c\) = \frac{2v}{3Hl}\(\frac{\p m_c}{\p z}n_c\) . \nonumber
\end{eqnarray}

\section{Effect of string/particle production on perturbations} \label{ppapp}

In \secref{ppsec}, we solved \eqref{eq:pert1} using a Greens' function method.  Here, we will solve it by another method that serves as a check and adds some intuition regarding the behavior of the solutions.  From \eqref{eq:pert1} and \eqref{deln}, the equation we wish to solve is
\be \label{eq:pertapp}
	\ddot{\del z} + \(3 H +  {1 \over 2 \sigma \gamma^3}{d \bar f \over d \dot z} \) \dot{\del z}
	- e^{-2 H t} \frac{\nab^2}{\gamma^2}  \, \del z
	= -{m_0^2 \,  \over   2 \sigma \gamma^{3}}  \sqrt{\bar n}  e^{-3 H t/2} \, .
\ee
Recalling that an overbar denotes time-average, and that $d \bar f/d \dot z$ is  constant, this equation describes a damped harmonic oscillator with a spring constant that decreases with time, and a constant driving force that also decreases---but more slowly than the spring constant.  The oscillator is underdamped at early times, but becomes overdamped at a certain critical time.  

{Switching to conformal time, we have}
\be
	\del z_{k}''  -\( {2 \over \tau} + {\lambda \over  \tau}\) \del z_{k}'+ {k^{2} \over \gamma^2}  \, \del z_{k}
	= {B \over \sqrt{- \tau H^{2}}} \, ,
\ee
where $\lambda \equiv  (d \bar f/ d \dot z)/2 H \sigma \gamma^3$, and $B \equiv -  m_0^2 \sqrt{H \bar n}/2 \sigma \gamma^3$ are dimensionless parameters related to the amount of string or particle production.

The general solution is
\be \label{fullsol} \begin{split}
\del z_{k} &= C_1 \tau ^{\nu} J_\nu \( {k \tau \over \gamma }  \) + C_2 \tau ^{\nu} Y_\nu \(\frac{k \tau }{\gamma } \)  -\frac{B}{H k \gamma ^2}2^{-\frac{7}{2}-\frac{\lambda}{2}}  \pi  \sqrt{-\tau }  \left(\frac{k \tau }{\gamma }\right)^{-\frac{1}{2}-\frac{\lambda}{2}}  \times \\ & 
\times \left(k^3 \tau ^3 \left(\frac{k \tau }{\gamma }\right)^\lambda Y_\nu\( \frac{k \tau }{\gamma }\) \Gamma\left(\frac{3}{4}\right) {\cal H}\left[\left\{\frac{3}{4}\right\},\left\{\frac{5+\lambda}{2},\frac{7}{4}\right\},-\frac{k^2 \tau ^2}{4 \gamma ^2}\right]+ \right. \\ &
\quad + \left. 2^{2 \nu} \gamma ^3 J_\nu\(\frac{k \tau }{\gamma }\) \text{Csc}\left[\nu \pi \right] \Gamma \left(-\frac{3}{4}-\frac{\lambda}{2}\right) \times
\right.\\ &
\quad \times \left. {\cal H}\left[\left\{-\frac{3}{4}-\frac{\lambda}{2}\right\},\left\{-\frac{1}{2}-\frac{\lambda}{2},\frac{1}{4}-\frac{\lambda}{2}\right\},-\frac{k^2 \tau ^2}{4 \gamma ^2}\right]-k^3 \tau ^3 \left(\frac{k \tau }{\gamma }\right)^\lambda \right. \times \\ & \left.
\quad \times J_\nu\( \frac{k \tau }{\gamma }\right) \text{Cot}\left[ \nu \pi \right] \Gamma\left(\frac{3}{4}\right) {\cal H}\left[\left\{\frac{3}{4}\right\},\left\{\frac{5+\lambda}{2},\frac{7}{4}\right\},-\frac{k^2 \tau ^2}{4 \gamma ^2}\right]\right),
\end{split} \ee
where $\cal H$ is the regularized hypergeometric function.

The late-time ($\tau \to 0$) limit of this expression is simple:
$$
\del z_{k} \to -C_2 \frac{2^{2 + \nu}  \Gamma\left( 1 + \nu\right)}{\nu \pi } \({\gamma \over k} \)^{\nu}. $$
To fix the coefficient $C_2$, we expand \eqref{fullsol} in the limit $\tau \to -\infty$, and match it to the Bunch-Davies state
$$
\delta z_{k} \to {i H \tau \over 2 \sqrt{\sigma \gamma^2 k} }e^{-i k \tau/\gamma}
$$
(see \eqref{delzk}) at $\tau_e = \gamma M/k$.  Solving gives
$$
C_2 = {1 \over 4} \sqrt{\pi } \( {k \over \gamma } \)^{\lambda/2} \left( e^{i \phi_1} \frac{{2}   H }{\sqrt{2 \sigma \gamma^3 }} \({H \over M}\)^{\lambda/2}+ e^{i \phi_2} {B \over H} \sqrt{\pi} 2^{-\lambda/2} \frac{ \Gamma\left(\frac{3}{4 }\right)}{\Gamma\left(\frac{7 + 2\lambda}{4}\right)}\right),
$$
where $\phi_{1,2}$ are real.  Because the phase of $B$ is random and uncorrelated with the first term, the power spectrum will be the sum in quadrature of these two terms, and therefore $\phi_{1,2}$ drop out of the result.  The late-time perturbation is
\be \label{deltaz}
\del z_{k} \to {\sqrt{2} \, \Gamma\(\nu + 1 \) \over k^{3/2}} \( e^{i \phi_1} {H \over \sqrt{\sigma}} {   2^{\nu} \over \nu \sqrt{ \pi } }\( {H \over M} \)^{\lambda/2} + e^{i \phi_2} {B\over H}  \frac{ \Gamma\left(\frac{3}{4 }\right)}{\Gamma\left(\nu + 1/4\right)} \),
\ee
in agreement with \eqref{eq:fullspectrum}.

The solutions \eqref{fullsol}have three regimes of interest (see \figref{pp}).  When $\lambda \gg 1$ the mode is more highly damped than is  the case  from Hubble friction alone, and therefore the amplitude of the homogeneous modes is exponentially supressed.  However, the source term $B$ may compensate.  When $\lambda \ll 1$, the damping from string/particle production is negligible compared to Hubble friction, and the homogeneous modes are undamped.  However, there are still two regimes:  $B \gg H^{2}/\sqrt{\sigma}$, in which case the fluctuations due to string/particle production are larger than those produced by de Sitter fluctuations, or $B \ll H^{2}/\sqrt{\sigma}$, in which case the contribution of string/particle production is negligible.

\begin{figure}
\begin{center}$
\begin{array}{c c c }
	\includegraphics[angle=0, width=0.32\textwidth]{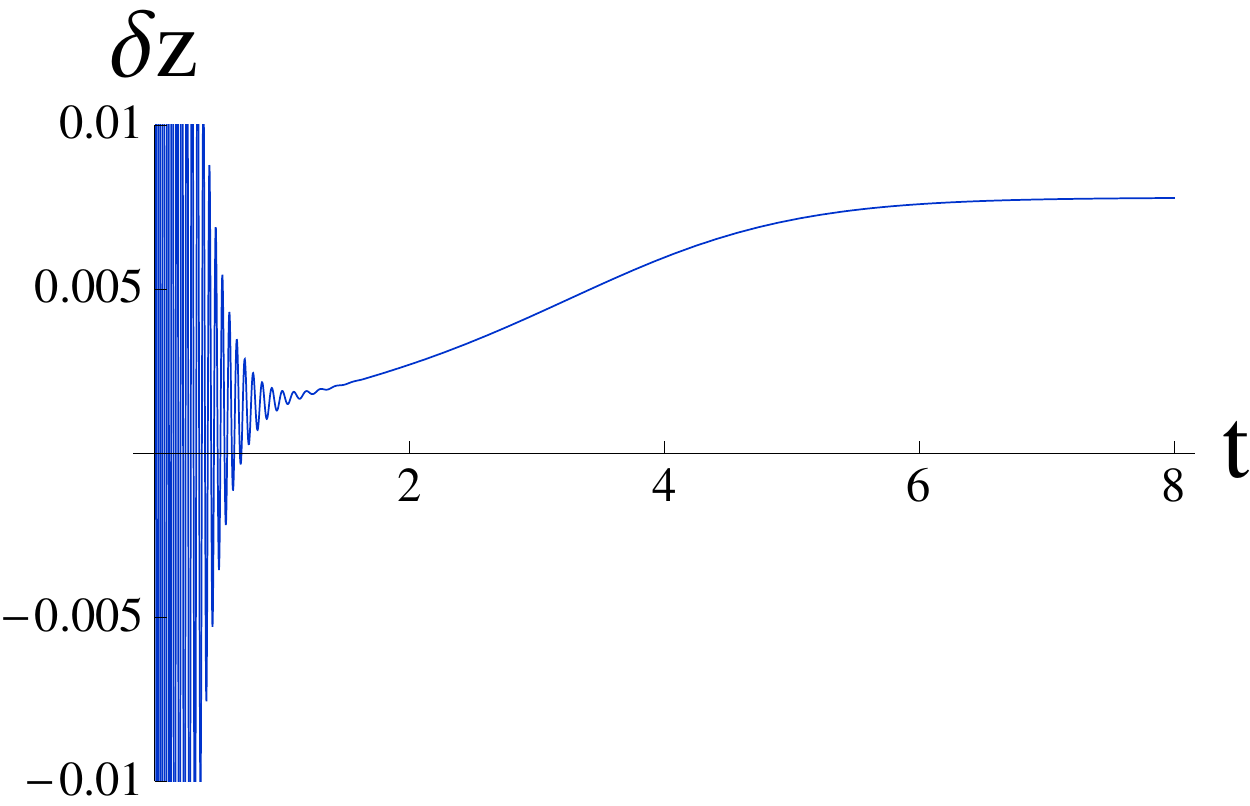} & \includegraphics[angle=0, width=0.32\textwidth]{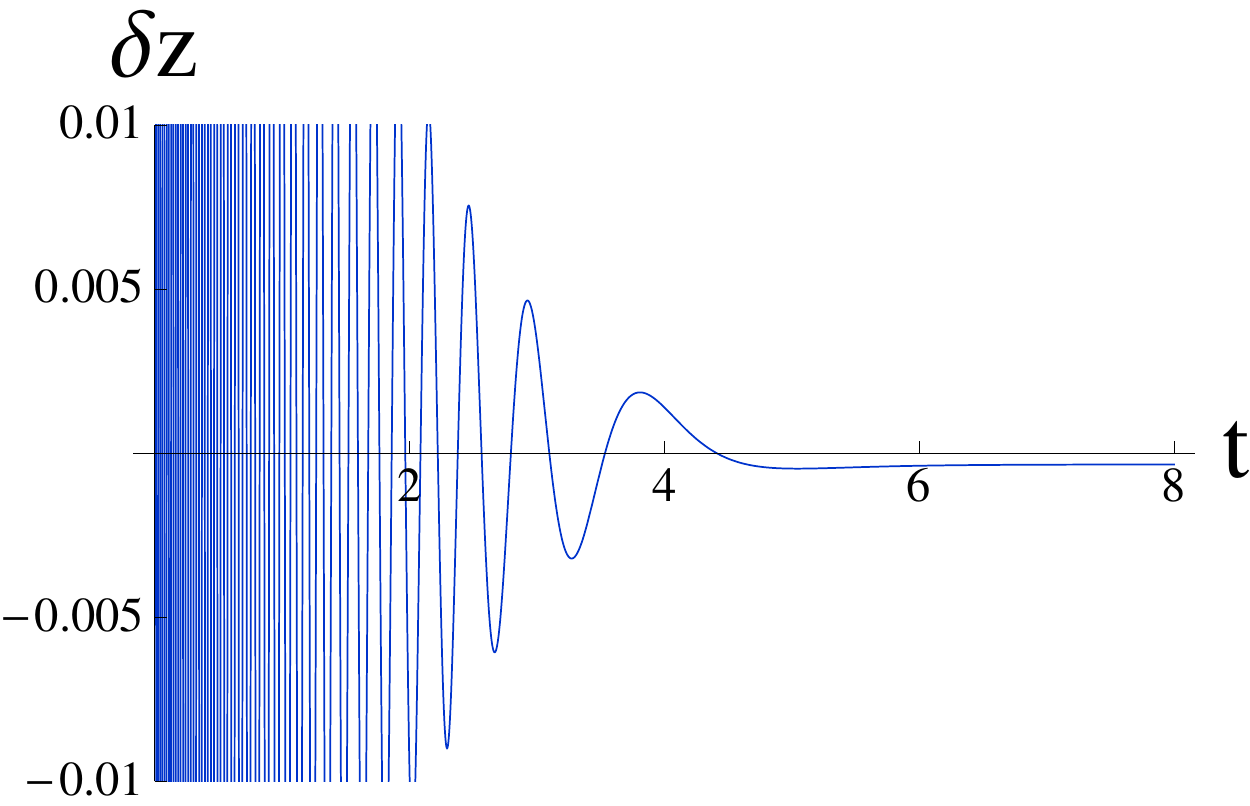} & \includegraphics[angle=0, width=0.32\textwidth]{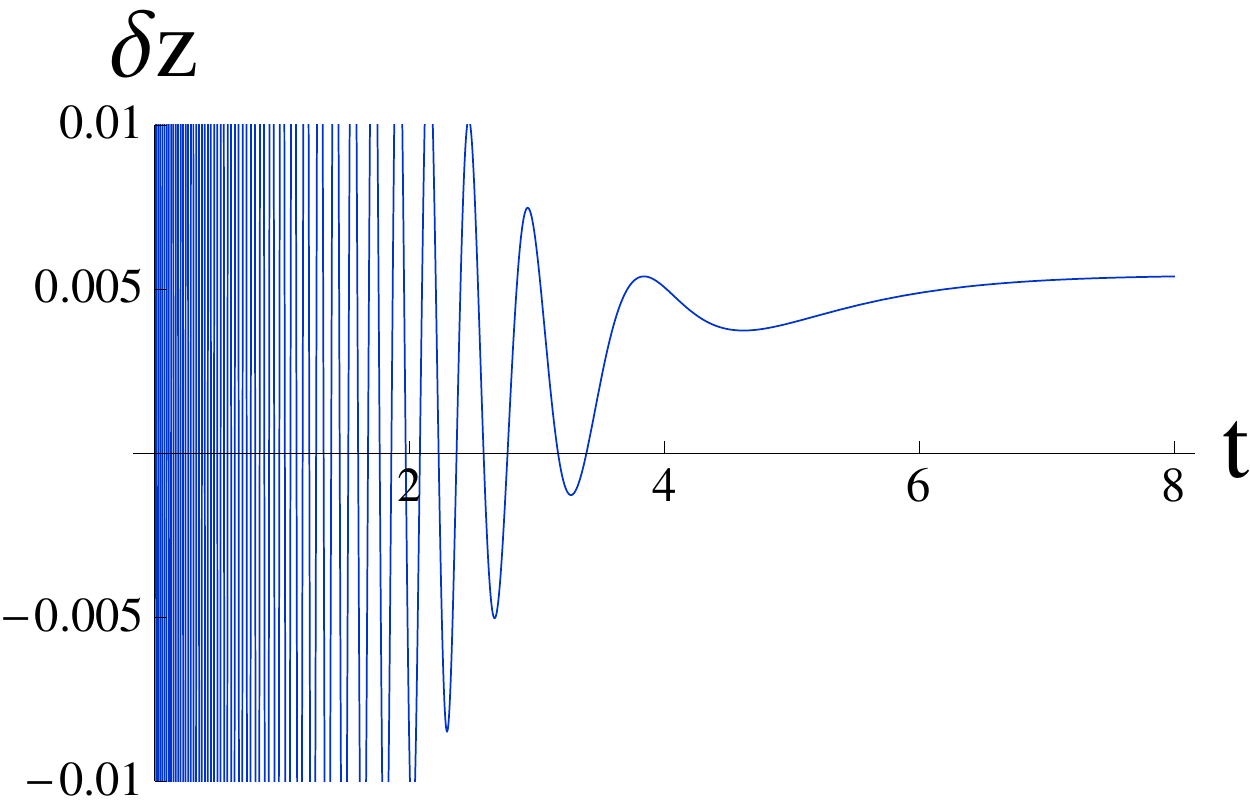} \\
\end{array}$
\end{center}
\caption{\label{pp} {Left panel: the behavior of the modes $\delta z$ in the presence of particle production, when the friction term $\lambda > 1$.   The additional friction damps the oscillations  rapidly, and after they become overdamped the source term $B$ pushes $\delta z$ to a non-zero value.  Center panel: $\lambda \ll 1$ and $B < H^{2}/\sqrt{\sigma}$.  Right panel: $\lambda \ll 1$ and $B > H^{2}/\sqrt{\sigma}$.}}
\end{figure}

\section{The amplitude of oscillations from string or particle production} \label{kickapp}

Here, we estimate the error made in time averaging the equation for the perturbations due to string or particle production.  The equation for the perturbation is
$$
 2 \gamma^3 \sigma \delta \ddot z + 6 H \sigma \delta \dot z \gamma + V''(z) \delta z+ \delta f = 0,
 $$
 where
\be \nonumber 
\begin{split}
	\delta f = & \delta \sum_{i} {\partial m_c \over \partial z} \, n_c \, e^{-3 H (t-t_{i})} \,  \Theta(t-t_{i}) = \sum_{i}e^{-3 H (t-t_{i})}  \left\{  \delta \dot z_i  \( {\partial^2 m_c \over \partial z \partial \dot z_i} \, n_c +		{ \partial m_c \over \partial z} \, {\partial n_c \over \partial \dot z_i}\) \Theta(t-t_{i}) + \right. \\
	& \left. \delta t_i {\partial m_c \over \partial z} \, n_c  \(3H \, \Theta(t-t_{i}) - \delta(t-t_{i}) \)  +  \delta n_c {\partial m_c \over \partial z} \,   \Theta(t-t_{i})\right\} .
\end{split}
\ee

When the number of collisions per Hubble time is large, one expects time averaging to be a good approximation.  But we may be interested in scenarios where the number of collisions per Hubble time is order 1.  

The terms proportional to $\delta \dot z$ are friction terms for the perturbation, but their average value will turn out to be much less than $H$ (at least in string theory), and therefore they have little effect with or without averaging.  The term $\delta t_i (3H \, \Theta(t-t_{i}) - \delta(t-t_{i}))$ oscillates and cancels after time averaging.   We can estimate the amplitude of the oscillations caused by these terms as follows.  The effect of the $\delta$ function is cancelled by the $\Theta$ function on average.  Therefore the amplitude of the oscillation is the ``kick" provided by the $\delta$ function.  The equation for the perturbation near the time of a collision will be approximately 
$$
2 \sigma \gamma^3 \delta \ddot z ={\partial m_c \over \partial z} \, n_c  \delta(t-t_{i}) \delta t_i ={\partial m_c \over \partial z} \, n_c  \delta(t-t_{i}) \delta z/\dot z.
$$
Integrating both sides gives
$$
\Delta \delta \dot z / \delta z =  {{\partial m_c \over \partial z} \, n_c / (\dot z 2 \sigma \gamma^3)},
$$
and therfore
$$
\Delta \delta z / \delta z= \Delta \delta \dot z/H \delta z = {{\partial m_c \over \partial z} \, n_c / (\dot z 2 H \sigma \gamma^3)}.
$$
This can be evaluated exactly given the underlying model.  To get a rough sense of its magnitude, if $n_c \sim m_s^3, \partial m_c/\partial z \sim m_s^2$, and $\sigma \sim m_s^4/g_s$, one obtains
$$
\Delta \delta z / \delta z \sim  {m_s g_s \over H \gamma^3}.
$$
This is very small in the models considered in \secref{stringsec}.

Lastly, the term proportional to $\delta n_{c}$ is a driving force that displaces the mode from its minimum at $\delta z = 0$.  
The combination of the $\Theta$ functions and the decaying exponentials make the sum oscillate around its average value, which is $ \approx \delta n_c {\partial m_c \over \partial z}/(3 H \Delta t)$, where $\Delta t \equiv t_{i} - t_{i-1}$.  The amplitude of the oscillations in the driving force is simply $\delta n_c {\partial m_c \over \partial z}$, and therefore the amplitude divided by the average is
$$
\Delta \delta n_c / \delta n_c \sim 3 H \Delta t.
$$
 The perturbation $\delta z$ due to string/particle production is  proportional to the driving force (see \eqref{deltaz}) when the driving force is constant.  Therefore oscillations in the driving force with frequency $\omega = 2 \pi/\Delta t \simleq H$ should introduce oscillations in $\delta z$ with the same amplitude.  Hence when $H \Delta t \sim 2 \pi$, the oscillations in the power spectrum due to string production are $\sim 1$.  

We have verified this conclusion numerically, and found that the amplitude of the oscillations is $10^{-2}$ when $H \Delta t \sim .5$, and fall off thereafter like a high power of $H \Delta t$.  In the models of \secref{stringsec}, this is the leading source of oscillations in the power spectrum.

\bibliography{Bibliography3}

\end{document}